\documentclass[amsmath,amsfonts,a4paper,11pt]{article}
\pdfoutput=1

\usepackage{jcappub} 
\usepackage{amssymb}
\usepackage{color}
\usepackage{graphicx,psfrag}
\usepackage{amsmath}
\usepackage{hyperref}
\usepackage{bbold}
\usepackage{braket}

\usepackage{natbib}

\usepackage[applemac]{inputenc}

\usepackage{soul}
\definecolor{dm}{cmyk}{.20, 0, .30, 0}
\definecolor{orange}{cmyk}{.0, .6, .90, 0}

\sethlcolor{dm}

\newcommand{\dd}{\mathrm{d}}

\newcommand{\epsilonH}{\ensuremath{\epsilon_{\rm H}}}
\newcommand{\Mpl}{\ensuremath{M_{\rm Pl}}}
\newcommand{\tic}{\ensuremath{\tau_{\rm IC}}}
\newcommand{\ti}{\ensuremath{\tau_{\rm i}}}
\newcommand{\hi}{\ensuremath{{\cal H}_{\rm i}}}
\newcommand{\h}{\ensuremath{{\cal H}}}
\newcommand{\hic}{\ensuremath{{\cal H}_{\rm IC}}}
\newcommand{\nic}{\ensuremath{n_{\rm IC}}}

\newcommand{\Vref}{\ensuremath{V_{\rm ref}}}
\newcommand{\phiref}{\ensuremath{\Delta \phi_{\rm ref}}}
\newcommand{\Href}{\ensuremath{H_{\rm ref}}}

\newcommand{\Vi}{\ensuremath{V}}
\newcommand{\phii}{\ensuremath{\Delta \phi}}
\newcommand{\Hi}{\ensuremath{H_{\rm i}}}
\newcommand{\ri}{\ensuremath{r}}

\newcommand{\Hic}{\ensuremath{H_{\rm IC}}}

\makeatletter
\DeclareFontEncoding{LS1}{}{}
\DeclareFontSubstitution{LS1}{stix}{m}{n}
\DeclareMathAlphabet{\mathcal}{LS1}{stixscr}{m}{n}
\makeatother

\def\be{\begin{equation}}
\def\ee{\end{equation}}
\def\bea{\begin{eqnarray}}
\def\eea{\end{eqnarray}}

\title{\boldmath{Inhomogeneous Initial  Data and Small-Field Inflation}}
\author{M.C.~David Marsh, John D. Barrow, and Chandrima Ganguly}

\affiliation{Department of Applied Mathematics and Theoretical Physics, University of Cambridge\\Wilberforce Road,  CB3 0WA, Cambridge, UK}

\emailAdd{m.c.d.marsh@damtp.cam.ac.uk}
\emailAdd{j.d.barrow@damtp.cam.ac.uk}
\emailAdd{c.ganguly@damtp.cam.ac.uk}

\abstract{We consider the robustness of small-field inflation in the presence of scalar field inhomogeneities. Previous numerical work has shown that if the scalar potential is flat only over a narrow interval, such as in commonly considered  inflection-point models, even small-amplitude inhomogeneities   present at the would-be onset of inflation at $\tau = \ti$ can disrupt the accelerated expansion. In this paper, we parametrise and evolve the inhomogeneities from an earlier time $\tic$ at which the initial  data were imprinted, and show that for a  broad range of inflationary and pre-inflationary models, inflection-point inflation withstands initial inhomogeneities. We consider three classes of perturbative pre-inflationary solutions (corresponding to energetic domination by the scalar field kinetic term, a relativistic fluid, and isotropic negative curvature), and two classes of exact solutions to Einstein's equations with large inhomogeneities (corresponding to a stiff fluid with cylindrical symmetry, and anisotropic negative curvature). We derive a stability condition that depends on the Hubble scales $H(\ti)$ and $H(\tic)$, and a few properties of the pre-inflationary cosmology. For initial  data imprinted at the Planck scale, the absence of an inhomogeneous initial  data problem for inflection-point inflation leads to a novel, lower limit on the tensor-to-scalar ratio.
  }

\begin{document}

\maketitle
\flushbottom

\section{Introduction}

\label{sec:intro} 
Two striking features of the Cosmic Microwave Background (CMB) radiation
are its average 
smoothness and the non-trivial correlations of its small anisotropies 
on scales significantly larger than the Hubble radius when it decoupled from
electrons. 
The approximate scale-invariance of the underlying primordial perturbations
cannot be explained by causal dynamics in the standard hot big bang model,
and would be a surprising outcome from any quantum gravitational initial
state. The success of the theory of inflation \cite{Guth:1980zm,
Linde:1981mu, Albrecht:1982wi} springs from its ability to explain these
observations within a simple theoretical framework that permits detailed
observational tests.

Inflation drives the universe towards a locally highly homogeneous state,
and small quantum fluctuations, stretched beyond the Hubble radius by
inflation, provide the candidate seeds for large-scale structure in the
universe \cite{Mukhanov:1981xt,Guth:1982ec,
Hawking:1982cz,Starobinsky:1982ee, Starobinsky:1983zz}. The generic
predictions of inflation are in excellent agreement with all current
observations \cite{Ade:2015xua}; however, much remains to be understood
about how and why inflation happened, and what might have prevented it.

A puzzling aspect about inflation is the question of how it got started: while the enormous expansion during this era smooths any pre-existing inhomogeneities, sufficiently large initial inhomogeneities can prevent the energy density from becoming dominated by the potential energy, and cause the expansion to remain decelerating. In the simplest estimates, the onset of the accelerated phase requires a homogeneous patch extending several (or even many) Hubble radii, which arguably suggests fine-tuning of the pre-inflationary initial state.
This is
the problem of inhomogeneous initial data for inflation.

The question of the robustness of inflation is important as it affects the
naturalness of the inflationary framework. For ardent critics of inflation,
the inhomogeneous initial data problem 
is interpreted as a serious challenge to the entire inflationary paradigm
(see e.g.~\cite{Ijjas:2013vea}). However, research by several groups over
the past few decades have shown that the simplest formulation of the initial
data problem can be misleading \cite{Starobinsky:1982mr, Wald:1983ky,
Barrow:1984zz, Albrecht:1984qt, Albrecht:1985yf, Albrecht:1986pi,
Jensen:1986nf, Jensen:1986vy, Barrow:1987ia, Barrow:1986yf, KurkiSuonio:1987pq, Feldman:1989hh, Brandenberger:1988ir, Goldwirth:1989pr,
Goldwirth:1989vz, Goldwirth:1991rj, Muller:1989rp, Kitada:1992uh, Calzetta:1992gv, Calzetta:1992bp,
Bruni:1994cv, Vachaspati:1998dy, Berera:2000xz, Easther:2014zga,
Carroll:2010aj, Guth:2013sya, Linde:2014nna, Berezhiani:2015ola,
East:2015ggf, Kleban:2016sqm, Clough:2016ymm, Brandenberger:2016uzh,
Linde:2017pwt, Clough:2017efm}, and inflation is less sensitive to
inhomogeneities than one might naively expect. Perhaps most importantly,
numerical simulations of  general relativity coupled to a scalar field with a
flat inflationary potential have shown that even initial configurations
dominated by inhomogeneities 
typically lead to inflation \cite{Albrecht:1985yf,
Albrecht:1986pi,KurkiSuonio:1987pq, Goldwirth:1989pr, Goldwirth:1989vz,
Goldwirth:1991rj,Easther:2014zga, East:2015ggf, Clough:2016ymm}, at least as
long as the amplitude of the inhomogeneities of the scalar field, $\delta
\phi $, is smaller than the width of the inflationary region of the
potential, $\Delta \phi $. For large-field models with $\Delta \phi >1$,%
\footnote{%
Throughout this paper, we set the reduced Planck mass to one: $\Mpl
=1/\sqrt{8\pi G}= 2.4\times 10^{18}\,\mathrm{GeV}=1$.} this solves the problem of
inhomogeneous initial data without requiring any smoothness of the initial
patch \cite{East:2015ggf}. Moreover, if chaotic inflation is realised, a single smooth Planck sized domain can result in inflation and lead to an eternal process of self-reproduction \cite{Linde:1986fd}.

For models with $\Delta \phi <1$, the solution to the inhomogeneous initial
data problem is less immediate. 
It can be avoided in models with non-trivial topology \cite{Cornish:1996st,
Coule:1999wg, Linde:2004nz, Linde:2014nna}, and ameliorated in inflationary
potentials with extended plateau regions \cite{East:2015ggf, Clough:2016ymm}
(which can be achieved, for example~through a non-trivial kinetic terms \cite{Kallosh:2013yoa, Linde:2017pwt}). Moreover, if the universe went through
phases of both high-scale and low-scale inflation, or got stuck in a false
metastable vacuum before the final period of inflation, the inhomogeneities
present at the onset of the final phase of inflation are expected to be
small (cf.~e.g.~\cite{Guth:2013sya, Brandenberger:2016uzh}).

Nevertheless, for commonly considered small-field models with $\Delta \phi
\ll 1$, the initial data problem can still appear quite severe. 
Examples of such models include low-scale potentials that are flat only near
an inflection point. Inflection-point models are becoming increasingly
popular as they are automatically consistent with observational upper limits
on the tensor-to-scalar ratio, and appear to admit comparatively simple
ultraviolet completions into a string theory (see e.g.~\cite{Kachru:2003sx,
Baumann:2007np, Baumann:2007ah,Agarwal:2011wm,McAllister:2012am,Dias:2012nf}, or \cite{Baumann:2014nda} for a recent review). Recently, reference \cite{Clough:2016ymm}, 
numerically studied the impact of inhomogeneities present at the potential
onset of inflation in small-field inflection-point models (see also \cite%
{East:2015ggf, Clough:2017efm}), and found that even highly sub-dominant gradient energy
densities of the scalar field 
can spoil inflation. 
At face value, these results may be taken to suggest that the simplest
inflection-point models suffer severely from the inhomogeneity problem. In
this paper, we show that such a conclusion would be premature.

The conformal time $\tic$ at which the initial data were imprinted
(e.g.~when four-dimensional general relativity first gave an appropriate
description of the dynamics) may in general have far preceded the 
onset of inflation at conformal time $\ti$. During the pre-inflationary era
between $\tic$ and $\ti$, the comoving Hubble radius grew, and the most
dangerous modes for disrupting inflation had wavelengths far longer than the
Hubble radius at $\tic$. The power spectrum of inhomogeneities at $\tic$ is
not in general expected to be scale-invariant, but should go to zero as the
wavelength goes to infinity.

In this paper, we parametrise the spectrum of inhomogeneities at $\tic$ and 
show that the initial data problem depends on four properties of the
scenario: \textit{i)} the energy scales $\Hic=H(\tic)$ and $\Hi=H(\ti)$, 
\textit{ii)} the amount of expansion between $\tic$ and $\ti$, \textit{iii)}
the spectral index of the initial inhomogeneities on super-horizon scales,
and \textit{iv)} the narrowing of the plateau region as the energy scale of
inflation is lowered. We then exemplify our results analytically by
considering pre-inflationary epochs dominated by either scalar-field kinetic
energy, a general relativistic fluid with equation-of-state $w$, or negative
curvature (which may be isotropic or anisotropic). Using these
models, we assess the severity of the inhomogeneous initial data problem.

We find qualitatively different scenarios depending on whether the expansion
is decelerating (e.g.~as for fluid domination), or displays constant
comoving Hubble parameter, ${\cal H=} Ha$, (as for negative
isotropic curvature domination), and, in the former case, if the fall-off of
the initial power spectrum on large scales is `steep' or `moderate'. This
leads us to three main results: 

\begin{itemize}
\item If the pre-inflationary expansion is decelerating and the initial data
are imprinted at the Planck scale, so that $\Hic=1$, the inhomogeneous
initial data problem only ever becomes relevant for very low energy 
models. For `moderate' initial power spectra, we derive a lower limit on the
tensor-to-scalar ratio $r$, above which one should not expect an initial
data problem. This limit implies that a large fraction of interesting
inflection-point models are robust against inhomogeneities, with the
detailed bound depending on a combination of parameters of the
pre-inflationary cosmology. For example, if the pre-inflationary epoch is
radiation dominated and the inhomogeneities at $\tic$ have $\delta \phi \sim 
\mathcal{O}(1)$ on the horizon scale and a spectral index of $\nic=3$ on
larger scales, we find the limit: $r>2.5\times 10^{-22}$. 
For `steep' initial spectra, no models are expected to have a problem with
inhomogeneities.

\item If the initial  data are imprinted at energies much below the
Planck scale with a `moderate' power-spectrum, the inhomogeneous initial
data problem becomes more severe. We show that this leads to a lower bound
on the initial energy scale, $\Hic\gtrsim 4\times 10^{-8}$, for models
robust against initial inhomogeneities. 
By contrast, models with `steep' initial power spectra can be robust against
inhomogeneities for smaller $\Hic$.

\item If the pre-inflationary dynamics is dominated by isotropic negative curvature so
that the comoving Hubble expansion rate is constant, there is
no inhomogeneous initial data problem for $\Hic\gtrsim 4\times 10^{-8}$,
independently of $\Hi$. 

\end{itemize}

We conclude that even the simplest inflection-point models with narrow
inflationary plateaux do not in general exhibit an inhomogeneous initial
data problem. 
While our results are derived for a particular class of small-field
inflection-point models (that directly generalise those considered in \cite%
{Clough:2016ymm}), we expect similar arguments to hold also for other
small-field scenarios. Moreover, 
our main results in \S\ref{sec:results} apply to pre-inflationary cosmologies in which the expansion is on average non-accelerating and the superhorizon scalar field inhomogeneities are not significantly sourced. We expect these conditions to be satisfied by many interesting classes of both perturbative and non-perturbative pre-inflationary cosmologies.

This paper is organised as follows: In \S\ref{sec:model}, we introduce
the family of inflection-point models that we  consider, review how inflationary models can fail due to inhomogeneities. We also parametrise their initial spectrum at $\tic$. In \S\ref{sec:preinflation}, we discuss the
pre-inflationary era and derive some simple but illuminating analytic
results for the evolution of perturbative inhomogeneities during inflation.
We also briefly discuss the cases of positive and anisotropic negative curvature. 
 In \S\ref{sec:results},
we analyse the severity of the problem of initial inhomogeneities, and draw
together the main results of this paper. We conclude by discussing possible
future directions in \S\ref{sec:concl}. In Appendix \ref{app:pert}, we provide details of the perturbative calculations in \S\ref{sec:preinflation}, and in Appendix \ref{app:nonlinear} we discuss solutions with cylindrical symmetry and anisotropic negative curvature  in more detail. 

\section{The initial data problem for inflection-point inflation}
\label{sec:model} 
In this section, we briefly review the class of inflationary models we consider and the wavelength-dependence of `dangerous' inhomogeneities. We also discuss and parametrise the initial spectrum of inhomogeneities, which will feature in our bounds derived in \S\ref{sec:results}. 

\subsection{Inflection-point inflation}
In order to investigate the dependence of the
inhomogeneous initial data problem on the energy scale $\Hi$, we 
construct a one-parameter family of models with various values for the
inflationary potential, $V_{0}$. These models directly generalise the
`typical small-field model' of reference \cite{Clough:2016ymm}.

\begin{figure}
 \centering
 \includegraphics[width=.7\textwidth]{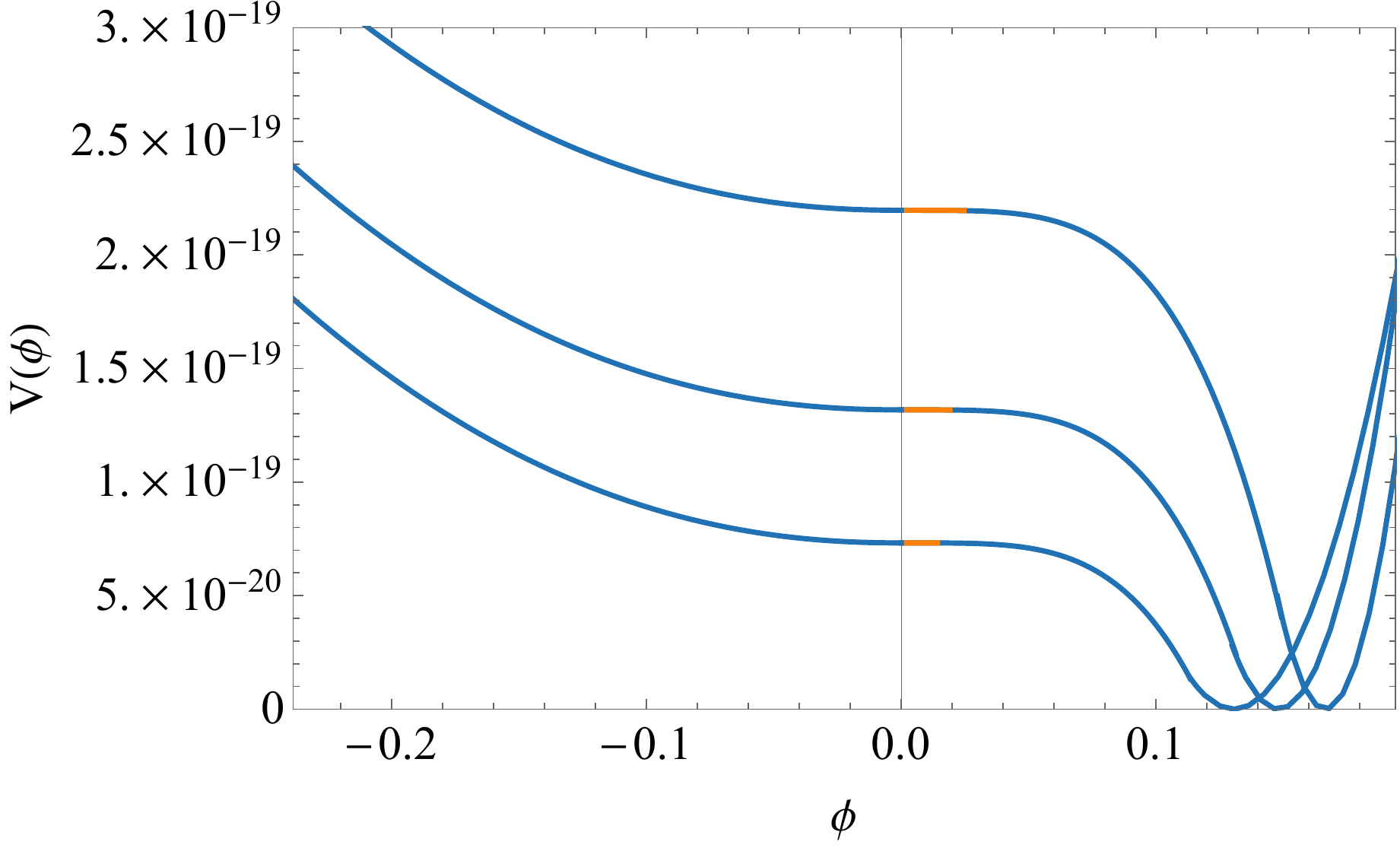}
 \caption{Examples of inflection-point potentials. Most of the inflationary expansion occurs when the field is in a narrow interval of length $\Delta \phi \sim {\cal O}( \mu^2)$, here marked in orange.}
 \label{fig:potentials}
 \end{figure}

The initial data problem is expected to be most severe in models with very
narrow inflationary plateaux, $\Delta \phi \ll 1$. In homogeneous cosmology,
the minimum required width of the inflationary plateau shrinks as $V_{0}$ is
decreased: the lower the energy scale, the narrower the plateau, cf.~Figure %
\ref{fig:potentials}. In this section, we briefly review the relevant
properties of inflection-point potentials and show why modes with $k\approx %
\hi=a_{\mathrm{i}}\Hi$ are the most dangerous for inhibiting inflation.

The models we consider have an inflection point at $\phi =0$ and for $0<\phi
\lesssim \mu $ are described by, 
\be
V(\phi )=V_{0}\left( 1-\left( \frac{
\phi }{\mu }\right) ^{4}\right) \thinspace\ . \label{eq:plateau} 
\ee 
For $%
\mu <1$, this potential supports small-field, slow-roll inflation. Clearly,
equation \eqref{eq:plateau} captures the plateau region adjacent to the
inflection point, and should be joined on either side by suitable smooth
extensions. 
In \S\ref{sec:results}, we consider both sharply rising and flat extensions of the potential to negative values, however, we note that  
the former class  are likely to suffer from the  (homogeneous)
`overshoot problem', as discussed in e.g.~\cite{Brustein:1992nk,
Dutta:2011fe} (see also \cite{Baumann:2007ah, McAllister:2012am} for a
discussion in the context of a multi-field string theory embedding of
inflection-point inflation). We will not address the overshoot problem here, but note that it motivates focussing on rather flat extensions of the plateau. 
%
In  this section, we assume that the average value of the field is close to the
inflection point when $H=\Hi$, with a small kinetic energy, so that the relevant part of the potential is
captured by equation \eqref{eq:plateau}.

The properties of this family of inflection-point models are well-known and
reviewed in e.g.~\cite{Lyth:1998xn}. 
If slow-roll inflation begins at $0<\phi _{\mathrm{min}}\ll \mu $, the
number of e-folds of inflation is well-approximated by, 
\be 
N = \int_{\phi _{
\textrm{min}}}^{\phi _{\textrm{max}}} \frac{1}{\sqrt{2\epsilon _{V}}} \dd%
\phi \approx \frac{1}{8} \frac{\mu ^{4}}{\phi _{\textrm{min}}^{2}}
\thinspace\ , 
\ee 
where $\epsilon _{V}=\tfrac{1}{2}(V^{\prime 2}/V^{2})$.
The spectral index of the curvature perturbations, evaluated around $\phi _{%
\mathrm{min}}$, is given by, 
\be
n_{s}-1 = -\frac{3}{N} \thinspace\ , 
\ee %
so that observational compatibility requires $N\approx 100$ e-folds of
inflationary expansion, independently of the energy scale of inflation.
Assuming that inflation ends soon after the moment when $|\eta
_{V}|=|V^{\prime \prime }/V|=1$, the distance traversed by the field during
inflation is approximately given by, 
\be 
\phi_{\rm{max}}- \phi_{\rm{min}}
\approx
{\cal O}(\mu^{2})
\, . \label{eq:Dphi} 
\ee 
Clearly, models with small field excursions have $\mu <\mathcal{O}(1)$.

Finally, the amplitude of the curvature perturbations generated from quantum
fluctuations during inflation scales like, 
\be
A_s \sim N^3 \frac{V_0}{\mu^4} \, . 
\label{eq:As}
\ee
Imposing the observationally inferred
normalisation of the CMB anisotropies then leads to a one-parameter family
of models.

Starting from an observational compatible reference model with $V_{0}=\Vref$
and the inflationary field-displacement $\Delta \phi =\phiref$, it follows
from equations \eqref{eq:Dphi} and \eqref{eq:As} that other observationally
compatible models in this family have field displacements $\phii$ and energy
scales $\Vi$ given by the scaling relation, 
\be
\left(
\frac{\phii}{\phiref}
\right)^2
=
\frac{V_0}{\Vref} = \left(\frac{\Hi}{\Href}\right)^2 \, .
\label{eq:scaling}
\ee
%
%


We  use the `typical small-field model' of \cite{Clough:2016ymm} for our
reference parameters:\footnote{%
These numerical values differ from those of \cite{Clough:2016ymm} which sets
the Planck mass, as opposed to the reduced Planck mass, to unity.} 
\begin{align}
\begin{array}{l l l}
\mu_{\rm ref} =0.12 \, ,
& \Vref= 7.3\times10^{-20} \, , ~~~\\
\Href = 1.6\times10^{-10} \, ,~~~
&
 \phiref = 5.0\times10^{-3} \,  ,
 \end{array}
 \label{eq:ref}
 \end{align}
where we have used the field excursion during all but the last e-fold of
inflation as our measure of $\Delta \phi $. For these parameters, the
tensor-to-scalar ratio is $r=2.6\times 10^{-12}$. 

Small-field models satisfy $\Delta \phi <1$. For this class of
inflection-point models, this translates into the limits $\Hi < 3.2\times
10^{-8}$ corresponding to $\ri < 10^{-7}$.



\subsection{Dangerous inhomogeneities}

\label{sec:danger} Some inhomogeneities are more dangerous to inflation than
others. Inflation is destabilised if, in some part of space, the scalar
field fluctuates towards the minimum and pre-maturely ends inflation in that
region, with gradients dragging the field in the rest of spacetime down
towards the minimum. However, gradients also have a stabilising effect \cite%
{East:2015ggf, Clough:2016ymm}, as we  now review.


For a potentially destabilising fluctuation to become energetically
favourable, the energy gain from the potential energy, $\Delta V$, must
overcome the gradient energy, $\tfrac{1}{2a^{2}}(\nabla \phi )^{2}$, that
will (at least initially) attempt to pull the scalar field fluctuation back
towards the plateau. Clearly then, for a fixed $k$-independent amplitude of
scalar field inhomogeneities, low-$k$ modes are more dangerous than high-$k$
modes. Inhomogeneities with $k<\hi$ can locally be viewed as
renormalisations of the homogeneous cosmology, and do not cause the entire
universe to collapse \cite{East:2015ggf}. Thus, the modes most dangerous for
inflation have $k\approx \hi=\Hi a_{\mathrm{i}}$. As we will discuss quantitatively below, reference \cite%
{Clough:2016ymm} found numerically that the probability 
of destabilisation decreased markedly for wavelengths a factor of two
smaller than the Hubble radius. These
results are consistent with earlier work \cite{East:2015ggf}.

For non scale-invariant inhomogeneities, the long-wavelength modes remain
the most dangerous as long as the potential energy gain is dominated by
linear or quadratic terms. If the potential energy gain becomes dominated by
cubic or higher order terms, it is possible for short-wavelength modes to
become dangerous: in the cubic case, this  only happens for rather steep
spectra of inhomogeneities at $\ti$, corresponding to a spectral index of $%
\gtrsim 5$. 

\begin{figure}
 \centering
 \includegraphics[width=.7\textwidth]{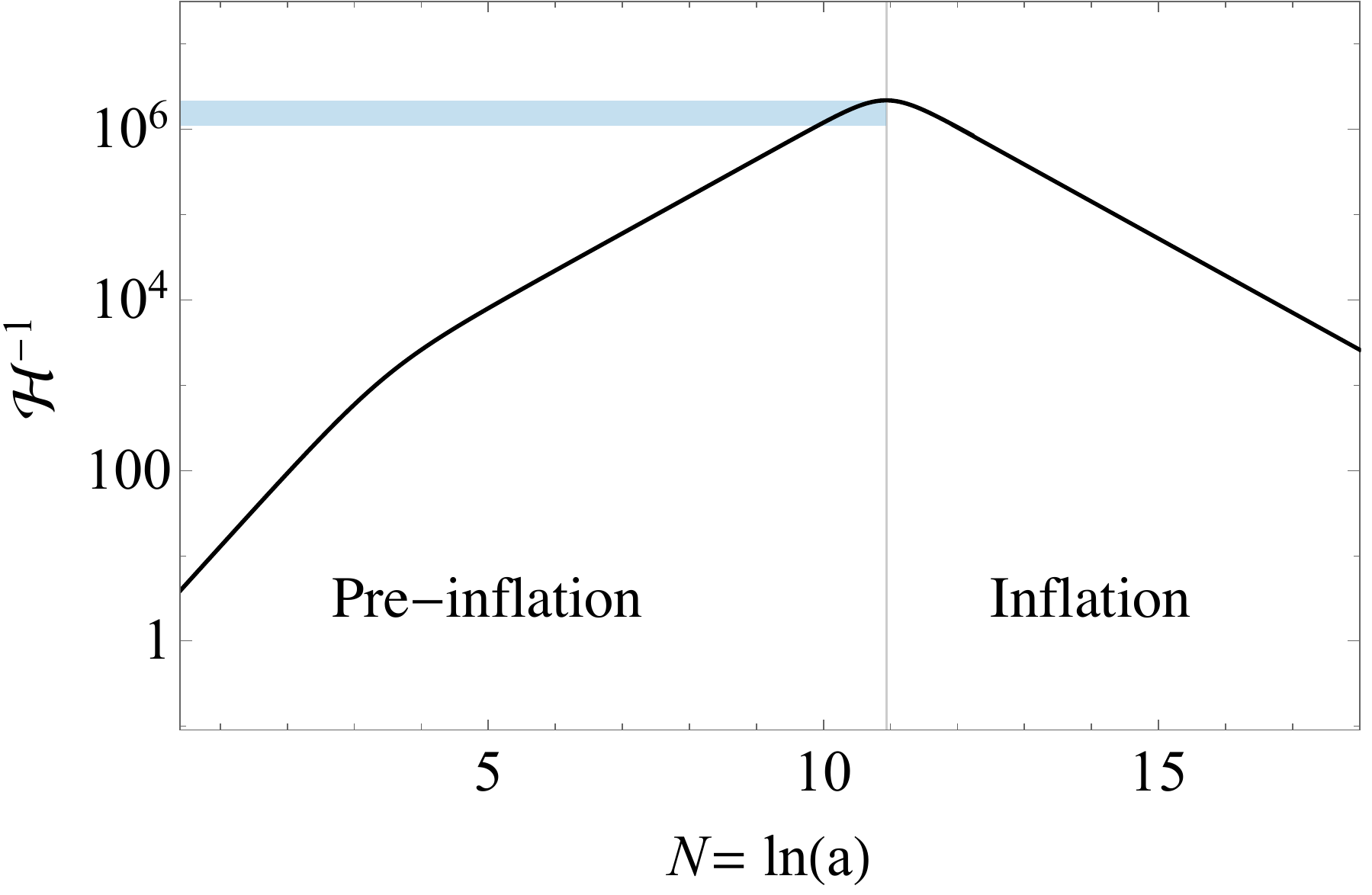}
 \caption{Evolution of the comoving Hubble radius, $\h^{-1}$, in a pre-inflationary epoch of decelerated expansion 
 driven by scalar field kinetic energy and radiation
 (solid) and during inflation (dashed). 
 Here $a(\tic)=1$. 
 The perturbations most dangerous for inflation (shaded blue) have $k^{-1} \gg \hic^{-1}$. }
 \label{fig:hubbleradius}
 \end{figure}

The sub-horizon evolution of high-$k$ inhomogeneities mitigates the risk
they pose for inflation: perturbative inhomogeneities decay like $\delta
\phi _{k}\sim 1/a$ (as we show in \S\ref{sec:perts}), and
non-perturbatively large inhomogeneities can trigger gravitational collapse
into black holes promptly after horizon entry \cite{Carr:2014pga}. The black holes are not expected to disrupt inflation \cite{Clough:2016ymm}.
For these reasons, we here focus on the limited range of `dangerous modes'
that have wave numbers in a small interval around $\hi$, cf.~Figure \ref{fig:hubbleradius}. This drastically simplifies the problem of inhomogeneous
initial data.

Inflation is robust against the effects of sufficiently small
inhomogeneities. 
The exact limit on their amplitude depends on the width of the inflationary plateau, the steepness of the potential beyond it, the admixture of wavelengths of the modes, and the criterion for robustness. We here adopt the criteria that a model is safe from inhomogeneities if it yields 60 or more e-folds of inflation. 
Extrapolating the numerical results of reference  \cite{Clough:2016ymm},\footnote{We are grateful to Eugene Lim, 
Josu Aurrekoetxea, Katy Clough, and Raphael Flauger
for discussions on this point.}
we express the corresponding bound on the total amplitude of scalar
field inhomogeneities with $k\approx \hi$ as a fraction $f$, of the width $\Delta \phi_{\rm ref}$:
\be
|\delta \phi_{ k \approx {\cal H}_i}| < f\, \Delta \phi_{\rm ref} \, .
\label{eq:cond1}
\ee
Here $\Delta \phi_{\rm ref}$ is as in equation \eqref{eq:ref}. We consider two values for the fraction $f$: if the scalar potential is flat for  negative values of the potential, $f =  1.6$, while if it raises sharply beyond the plateau, $f=0.17$. The latter type of potential suffers from the overshoot problem already for homogeneous cosmology \cite{Dutta:2011fe}, and the reduction of $f$ is directly related to this problem; inhomogeneities can pick up excess kinetic energy by fluctuating to negative values. We expect that potentials without an overshoot problem give $f > 0.17$.

These values for $f$ are obtained from numerical simulations in which all inhomogeneities were concentrated in a superposition of three modes with comoving wavelengths  $\hi^{-1}$. Reference \cite{Clough:2016ymm} also found that if, in addition, modes with half the wavelength were included, the corresponding values were increased to $f=2.6$ and $f=0.21$ for extended flat and steep potentials, respectively.
This suggests that $f$ is a $k$-dependent function, consistent with the argument that long wavelength modes are the most dangerous. Due to the
scarcity of numerical data,  we will not
attempt to model $f(k)$ here.
 Moreover, we will not consider other sources
of inhomogeneity, e.g.~those in tensor modes, that only indirectly impact
the stability of inflation \cite{Clough:2017efm}.

Using the scaling equation \eqref{eq:scaling}, the condition for stability %
\eqref{eq:cond1} can be extended to other energy scales in our class of
inflection-point models: 
\be
|\delta \phi_{ k \approx {\cal H}_i}| < f \Delta \phi 
=   f \phiref \left(\frac{H_i}{\Href}\right) \, .
\label{eq:bound1}
\ee

 In \S\ref{sec:results} we  use this inequality to
derive a bound on $\Hi$ and $\Hic$ from the absence of an inhomogeneous
initial data problem.

\subsection{Parametrisation of the initial inhomogeneities}

\label{sec:nic}

Presumably, inhomogeneities present at $\tic$ were  fashioned by
quantum gravitational dynamics, about which little is known.\footnote{%
In the context of string theory -- the leading candidate theory of quantum
gravity -- the question of inflationary initial  data is likely to
involve the properties of the effective theories at energies below the
compactification scale, and the state of the universe at energies at or
above the string scale. While substantial, yet partial, progress have been
made on the former issue, the latter still remains a significant challenge
(see e.g.~\cite{Baumann:2014nda} for a review).} In lieu of a complete
theory of the initial  data, we here merely parametrise the statistical
distribution of initial inhomogeneities, and briefly highlight the most
relevant properties of their spectrum. 

We assume that the inhomogeneities present at $\tic$ are classical and
statistically homogeneous, isotropic and Gaussian. We define the power
spectrum of inhomogeneities at $\tic$ by, 
\be
\langle \delta \phi_{\bf k}(\tic) \delta \phi_{\bf k'}(\tic) \rangle  = \frac{2\pi^2}{k^3} \delta^{(3)}({\bf k} + {\bf k}') P_{\delta \phi(\tic)}(k) \, .
\ee
In general, $P_{\delta \phi (\tic)}(k)$ can be a
complicated function, however, we expect the power spectrum to be suppressed
on wavelengths greater than the largest dynamically relevant length scale.
In particular, as the wavelength of the modes goes to infinity (and $%
k\rightarrow 0$), we expect $P_{\delta \phi (\tic)}(k)\rightarrow 0$. %
For modes with $k<\hic$, it may be appropriate to describe the
power-spectrum by a simple power-law,\footnote{%
If the dynamics responsible for the initial inhomogeneities includes
multiple length scales, the power-law parametrisation should be appropriate
for wavelengths larger than the largest dynamical scale, which may be longer
than $\hic^{-1}$. We do not discuss this modified scenario here, but our
analysis can be straightforwardly applied also to this scenario.} 
\be
P_{\delta \phi(\tic)}(k\, |\, k < \hic)  = A \left( \frac{k}{\hic}\right)^{\nic-1} \, ,
\label{eq:nic}
\ee
 in terms of which the assumption of $P_{\delta \phi (\tic%
)}(0)=0$ gives $\nic>1$. Reasonably, this is consistent with a finite total
power in modes with $k<\hic$. Here $A$ sets the amplitude of inhomogeneities with $k=\hic$. The energy density contribution from these modes is $\rho_{\delta \phi(\tic)|_{k=\hic}} = H^2 A$, which is clearly substantial for $A=1$.  

The spectral index of the initial inhomogeneities is an important
parameter in our analysis, and the condition $\nic>1$ can be motivated by
several additional arguments. Obviously, were we to assume that classical
cosmology with decelerating expansion would hold up until the singularity at 
$\tau =0$, no causal mechanism could generate perturbations with comoving
wavelengths $\gg \hic^{-1}$, and there would be no power in modes with $k\ll %
\hic$. Moreover, quantum fluctuations can lead to suppressed perturbations
on scales larger than the Hubble radius, as we  now show.

Neglecting gravitational back-reaction around a homogeneous FRW background
with constant $\epsilonH$, a massless scalar field perturbation $\varphi
_{k}=\delta \phi _{k}/a$ satisfies the linear-order equation, 
\be
\frac{\dd \varphi_{ k}}{\dd \tau^2} + \left( k^2 - \frac{c}{\tau^2} \right) \varphi_{ k} = 0 \, .
\label{eq:flatspaceeqn}
\ee 
 Here, we have used
that $\h=\tfrac{1}{\epsilonH-1}\frac{1}{\tau },$ and introduced the
constant, 
\be
c = \frac{2 - \epsilon_H}{(\epsilon_H -1)^2} \, ,
\ee 
 for $\epsilonH\neq 1$. In the case where the background
geometry is de Sitter spacetime, we have $c=2$ and the solutions to equation %
\eqref{eq:flatspaceeqn} include the familiar Bunch-Davies wavefunction, 
\be
\varphi_{ k} = \frac{e^{i k \tau}}{\sqrt{2k}} \frac{(k\tau - i)}{k\tau} \, .
\ee
In this case, long-wavelength modes have a power spectrum, %
\be
P_{\varphi} = \frac{k^3}{2\pi^2} \left| \varphi_k \right|^2 \sim \frac{1}{\tau^2} \, ,
\ee
independently of $k$ (for sufficiently small $k$) so that the power spectrum
of $\phi $ is scale-invariant: $P_{\delta \phi }\sim H^{2}$ \cite%
{Bunch:1978yq}. This is no longer the case in a more general FRW background
with constant $\epsilon _{H}>1$. Specialising to the case in which the
energy density is dominated by a relativistic fluid with equation-of-state
parameter $0<w<1$, we have $3/2<\epsilon _{H}<3$ and consequently $-\tfrac{1%
}{4}<c<2$. The solution to equation \eqref{eq:flatspaceeqn} is now given by
a combination of Bessel functions, 
\be
\varphi_k(\tau) = c_1 \sqrt{\tau} J_{\nu}(k\tau) + c_2 \sqrt{\tau} Y_{\nu}(k\tau) \, ,
\ee
with 
\be
\nu = \tfrac{1}{2}\sqrt{1+4c} = \frac{1}{2} \frac{|3-\epsilon_H|}{| \epsilon_H -1|}\, .
\ee
 The slowest
fall-off in the long-wavelength limit $k\ll \tau $ is obtained from the
`Bunch-Davies-like' solutions that scale as,
\be
\varphi_k(\tau) \sim \frac{1}{k^{\nu} \tau^{\nu -1/2}} \, . 
\ee  
The spectral index for the perturbations of $\phi$ is then given by,
\be
\nic -1 = 3 - 2\nu = 3-  \frac{|3-\epsilon_H|}{| \epsilon_H -1|} > 0\, .
\ee
 In the
entire range $0<w<1$, we have $\nic>1$ and thus a suppression of the power
in modes with wavelength far longer than the Hubble radius. For example, a
radiation-dominated period corresponds to $\nic=3$.

We note in closing that since the appropriate theory of inflationary initial
 data is unknown, arguments about its properties are necessarily
heuristic. 
In this section we have pointed out that scale invariance of the
inhomogeneities is certainly not guaranteed, and should perhaps not be
expected.

\section{The pre-inflationary epoch}
\label{sec:preinf}

\label{sec:preinflation} On general grounds, there is no reason to expect
that the time $\tic$ should coincide with the onset of inflation at $\ti$.
This means that inflation was preceded by a pre-inflationary era of
non-accelerated expansion.\footnote{%
More exotically, it is possible that the pre-inflationary era involved a
contracting phase. We will not discuss this scenario here.} Since inflation
by construction is very efficient at erasing any traces of the
pre-inflationary state, not much is known about this epoch.\emph{\ }

Our main results in \S\ref{sec:results} depend only on a few
properties of the pre-inflationary era, and apply to both small and large
inhomogeneities. For concreteness and to be able to make analytic progress,
in this section we parametrise the pre-inflationary spacetime by a
homogeneous and isotropic Friedmann-Robertson-Walker (FRW) metric, and treat
the inhomogeneities as linear-order perturbations. This approach has some
obvious limitations, but also several benefits. While the full problem of
determining how inhomogeneities affect the duration of inflation is clearly
non-linear, the \emph{pre-inflationary} evolution of small perturbations
with $\delta \phi \approx \Delta \phi \ll 1$ can still be well-described by
perturbation theory. For small-field models with a very narrow plateau
region, such small perturbations may suffice to destabilise inflation (at
which point the perturbative treatment ceases to be valid).

Moreover, as noted in \S\ref{sec:danger}, the most dangerous modes for
inflation had wavelengths far larger than the Hubble radius during most of
the pre-inflationary era. These modes are governed by a coarse-grained
effective theory obtained by integrating out short-wavelength
inhomogeneities. This results in a prescription similar to the `stochastic
inflation' framework \cite{Starobinsky:1986fx} (see also \cite{Moss:2016uix}), and includes a small
sourcing induced by short-wavelength inhomogeneities. Our
bounds derived in \S\ref{sec:results} applies to  pre-inflationary cosmologies in which this sourcing is negligible. In all models that  we study analytically in this section, the superhorizon modes are constant. 

This section is organised as follows: in \S\ref{sec:bkg}, we review the
salient features of homogeneous pre-inflationary cosmology, and in \S%
\ref{sec:perts} we discuss the evolution of perturbative inhomogeneities in
the three classes of pre-inflationary scenarios that we consider. 
In \S\ref{sec:poscurv} and \S\ref{sec:anisocur}  we respectively review the cases of positive and anisotropic negative curvature. 
More details of the perturbative calculations are deferred to Appendix \ref{app:pert}, and some special, non-linear solutions are further discussed in Appendix \ref{app:nonlinear}. 








\subsection{Homogeneous pre-inflationary cosmology}

\label{sec:bkg} 

In this section, we discuss the homogeneous limit of the pre-inflationary
cosmology and derive simple expressions for the duration of the
pre-inflationary era and the growth of the comoving Hubble radius. The
Friedmann-Robertson-Walker (FRW) metric is given by, 
 \be
 \dd s^2 = a^2(\tau) \left( - \dd \tau^2 + \gamma_{ij} \dd x^i \dd x^j\right) \, ,
 \ee
 with the spatial components,
 \be
 \gamma_{ij} =  \frac{\delta_{ij}}{\left(1+ \frac{\cal K}{4} (x^2+y^2+z^2)\right)^2
 } \, .
 \ee
We
 set $a(\tic)=1$ so the (isotropic) curvature parameter $\mathcal{K}$ is in general
not an integer. 
We here consider  three-dimensional
spaces that are flat or open with $\mathcal{K}=0$ or $\mathcal{K}<0$, and discuss the case of positive curvature in \S\ref{sec:poscurv}. A
useful parameter is, 
\be
\epsilon_H = - \frac{\frac{\dd H}{\dd t}}{H^2}= 1-\frac{{\cal H}_{\tau}}{{\cal H}^2} \, ,
\ee
 which,
in the pre-inflationary era of non-accelerated expansion, satisfies $%
\epsilon _{H}\geq 1$. Here, and subsequently, we denote derivatives with
respect to conformal time with the sub-script $\tau $. 

We consider universes containing a homogeneous inflaton field, $\bar{\phi}$,
with potential $V(\bar{\phi})$ and a perfect, isentropic fluid with energy
density $\rho _{f}(\tau )$ and pressure $p_{f}=w\rho _{f}$. In the
homogeneous limit, the Einstein equations are given by, 
\bea
 3 ({\cal H}^2 + {\cal K}) &=& \frac{1}{2} \bar \phi_{\tau}^2 + a^2 V(\bar \phi) + a^2 \rho_f  \, , 
 \label{eq:EE1}
 \\
 -( 2 {\cal H}_{\tau}+ {\cal H}^2 + {\cal K}) &=&  \frac{1}{2} \bar \phi_{\tau}^2 - a^2 V(\bar \phi) + a^2 p_f \, .
 \label{eq:EE2}
 \eea
 The Klein-Gordon equation and momentum conservation equation for the fluid are given by,
 \be
\bar \phi_{\tau \tau} + 2 {\cal H} \bar \phi_{\tau} + a^2 V'(\bar \phi) = 0\, , ~~~
 (\rho_{f})_{\tau} + 3 {\cal H}(\rho_f + p_f) =0
 \, .
 \ee

We  derive analytic results for three classes of cosmological
backgrounds: 
\begin{align}
\begin{array}{l l l r l}
i) & {\rm Kinetic~energy~domination:} & ~~~~~& \tfrac{1}{2 a^2} \bar \phi_{\tau}^2 ~&\gg {\rm max}\left( V(\bar \phi),~  \rho_f,-3 \tfrac{\cal K}{a^2}  \right) \\
ii) &{\rm Fluid~domination:} & ~~~~~&  \rho_f ~&\gg {\rm max}\left( \tfrac{1}{2 a^2} \bar \phi_{\tau}^2, ~ V(\bar \phi),-3\tfrac{\cal K}{a^2}  \right) \\
iii)~ & {\rm Curvature~domination:} & ~~~~~&  -3\tfrac{\cal K}{a^2}   ~&\gg {\rm max}\left( \tfrac{1}{2a^2} \bar \phi_{\tau}^2,~ V(\bar \phi), ~ \rho_f\right) \, .
\end{array}
\label{eq:classes} 
\end{align}

During kinetic energy domination
(considered also in \cite{Contaldi:2003zv, Handley:2014bqa}), $\epsilonH=3,$ and the energy density
decreases as $\rho =\rho (\tic)/a^{6}$. During fluid domination
(considered recently in \cite{Bahrami:2015bva}),\footnote{%
In highly inhomogeneous universes dominated by scalar field gradient energy
with wavelengths smaller than the Hubble radius (defined by spatial
averaging), the average energy density decreases approximately like
radiation, $w=1/3$ \cite{East:2015ggf}.} $\epsilonH=\tfrac{3}{2}(1+w)$ and $%
\rho (\tau )=\rho (\tic)/a^{3(1+w)}$. For isotropic curvature
domination (considered e.g.~in \cite{Freivogel:2005vv}), $\epsilonH=1$ and the isotropic curvature contribution decays
like $-\mathcal{K}/a^{2}$. In general, the energy density at $\tic$ may
receive contributions from multiple sources. The different decay rates of
the energy density then often lead to a separation of the pre-inflationary
era into distinct epochs during which one source dominates.

The total amount of expansion during a pre-inflationary era lasting from $\tic$ (when $H=\Hic$) to $\ti$ (when $H=\Hi$) is given by, 
\be
\frac{a(\ti)}{a(\tic)} = \left( \frac{\Hic}{\Hi}\right)^{\tfrac{2}{3(1+w)}} \, , 
\ee
 where $w=1$ corresponds to kinetic energy domination like an
`ultra-stiff' fluid, and $w=-1/3$ also corresponds to curvature domination.
For $w>-1/3$, the comoving Hubble radius, $\mathcal{H}^{-1}$, grows during
the pre-inflationary era by, 
\be
\frac{\hi^{-1}}{\hic^{-1}} = \left(
\frac{\Hic}{\Hi}
\right)^{\frac{1+3w}{3(1+w)}} \, .
\label{eq:hratio}
\ee
This is a
substantial growth for natural values of the parameters: for example, a
radiation-dominated pre-inflationary era lasting from the Planck scale ($\Hic%
=1$) to the energy-scale of our reference model (cf.~equation \eqref{eq:ref}%
) generates an expansion and growth of the comoving Hubble radius of $a(\ti%
)/a(\tic)=\hi^{-1}/\hic^{-1}=\mathrm{exp}(11.3)=8.0\times 10^{4}$. Moreover,
the energy scale of inflation may be much lower than that of the reference
model. Successful primordial nucleosynthesis and thermalisation of the
neutrinos requires a hot big bang cosmology with $T\gtrsim 4~\mathrm{MeV}$ 
\cite{deSalas:2015glj}, which in the extreme case of instant reheating
immediately following the end of inflation is consistent with $\Hi\approx
10^{-42}$ (corresponding to $r\approx 10^{-76}$). In this case a
radiation-dominated pre-inflationary era from the Planck scale would
generate 48 e-folds of expansion.

In the kinetic-energy dominated era, the homogeneous inflaton satisfies the
equation $\bar{\phi}^{\prime \prime }=0$, where prime denotes a derivative
with respect to the number of e-foldings: $X^{\prime }=\dd X/\dd N=\dd X/(\h%
\dd\tau )$. The speed of the field is then constant, with $\bar{\phi}%
^{\prime }=\pm \sqrt{6}$. In the other scenarios of equation %
\eqref{eq:classes}, the additional Hubble friction from the fluid or the
curvature term leads to the equation of motion, 
\be
\bar \phi'' + \frac{3}{2}(1-w)\bar \phi' =0 \, ,
\label{eq:bkgspeed}
\ee 
and a rapid slow-down of the background scalar field.%
\footnote{%
This effect ameliorates the overshoot problem for small-field inflation.} 

The growth of the co-moving Hubble radius, cf.~equation \eqref{eq:hratio}, is not unique to homogeneous FRW universes. In \S\ref{sec:anisocur} we show that similar results hold also for universes dominated by negative, anisotropic spatial curvature. 

\subsection{Perturbative pre-inflationary inhomogeneities}

\label{sec:perts} It is instructive to examine the pre-inflationary
evolution of small inhomogeneities in the three classes of cosmologies of
equation \ref{eq:classes}. To do so, we  consider the linear scalar
perturbations in conformal-Newtonian gauge. The line element is given by, 
\be
\dd s^2=
 a^2(\tau) \left(
-(1+2\Phi) \dd \tau^2 +   (1- 2 \Psi) \gamma_{ij} \dd x^i \dd x^j
\right) \, .
\ee
The benefit of this gauge is that the potentials $\Phi
,~\Psi $ agree with the gauge invariant Bardeen potentials and the scalar
field perturbation $\delta \phi (\tau ,\mathbf{x})=\phi (\tau ,\mathbf{x})-%
\bar{\phi}(\tau )$ directly corresponds to the gauge invariant scalar field
perturbation \cite{Mukhanov:1990me}. 
We assume that gravitational waves can be neglected. Inhomogeneous gravitational waves do not by themselves shorten the duration of inflation, and pre-inflationary cosmologies involving both tensor and scalar inhomogeneities was recently studied in 
 \cite%
{Clough:2017efm}. In the absence of anisotropic stress, the $i\neq j$
components of the Einstein equation gives, 
\be
 \Phi =  \Psi  \, ,
\ee
and we will henceforth only use the Newtonian potential, $\Phi $. The remaining Einstein equations are then given by (for reviews, see e.g.~\cite{Mukhanov:1990me, Malik:2008im}), 
\bea
\nabla^2 \Phi - 3 {\cal H} \Phi_{\tau} - 3({\cal H}^2 -{\cal K})\Phi &=& \frac{1}{2} \left(
a^2 V'(\bar \phi) \delta \phi + \bar \phi_{\tau} \delta \phi_{\tau} - \Phi \bar \phi_{\tau}^2 + a^2 \delta \rho_f
\right) \, , 
\label{eq:perts1}
\\
 \partial_i \left( {\cal H} \Phi + \Phi_{\tau} \right) &=& \frac{1}{2} \left(\bar \phi_{\tau} \partial_i \delta \phi - a^2(1+w) \bar \rho_f v_i  \right) \, , 
 \label{eq:perts2}
 \\
 \Phi_{\tau \tau} + 3 {\cal H} \Phi_{\tau} + (  2 {\cal H}_{\tau} + {\cal H}^2  - {\cal K})\Phi &=& \frac{1}{2} \left(
 -\Phi \bar \phi_{\tau}^2 
 + \bar \phi_{\tau} \delta \phi_{\tau} - a^2 V'(\bar \phi) \delta \phi + a^2 w \delta \rho_f 
 \right) \, .
 \label{eq:perts3}
\eea 
Here $v_i = \partial_i v$ for the fluid's velocity potential $v$. The leading-order Klein-Gordon equation for the field perturbation is given by,
\be
\delta \phi_{\tau \tau} + 2{\cal H} \delta \phi_{\tau} - \nabla^2 \delta \phi -4 \Phi_{\tau} \bar \phi_{\tau} + 2a^2 \Phi V'(\bar \phi) + a^2 V_{\phi \phi}(\bar \phi) \delta \phi = 0 \, .
\ee

We now state the results of this perturbative analysis; more details can be found in Appendix \ref{app:pert}. For all three classes of pre-inflationary cosmologies 
%
%
of equation \eqref{eq:classes},
%
we find that the amplitude of scalar field inhomogeneities stays constant on super-horizon scales, and undergo pressure-damped  oscillatory decay on sub-horizon scales.
In all three cases the amplitude of the scalar field inhomogeneities evolve like,
\bea
\delta \phi_{\bf k} \sim \left\{
\begin{array}{l c l}
{\rm constant} &~~~& k\ll {\cal H}\, , \\
e^{-N} &~~~ & k \gg {\cal H} \, .
\end{array}
\right.
\label{eq:decayrate}
\eea

In Appendix \ref{sec:Liang}, we review how the result of \cite{Liang} implies  that equation \eqref{eq:decayrate} also holds in the case of large, non-linear inhomogeneities with cylindrical symmetry.

\subsection{Positive curvatures}
\label{sec:poscurv}

Inflation cannot occur in regions that collapse before inflation can
commence. Such regions behave rather like local closed universes. Closed
universes are those with compact space sections. In the FRW case this
requires positive 3-curvature but in the most general closed universes (for
example those of Bianchi type IX) the curvature can change sign with time
and is mostly negative during any period of chaotic `mixmaster' dynamics: it
only becomes positive when the dynamics are close to isotropy. The fate of
critically overdense regions in the pre-inflationary period mirrors that of
the fate of closed universes. The most overdense regions may collapse to
form primordial black holes after they enter the horizon \cite{Carr:2014pga}, and then evaporate
primarily into massless and relativistic particles by the Hawking effect.
However, in any period of expansion that is dominated by the kinetic energy
of a scalar field the corresponding Jeans length equals the horizon size and
there is little scope for overdensities to collapse into black holes before
they can be supported by pressure. Therefore the most pronounced
overdensities will be filtered out by gravitational collapse. They can
collapse before inflation can begin and leave the remaining lower-density
regions to undergo inflation. Thus, in an inhomogeneous chaotic inflationary
scenario, only the lower density regions that avoid premature gravitational
collapse will be able to inflate and become candidate regions to contain our
visible universe.

Another scenario might ensue if overdense regions collapse and bounce
through a sequences of growing oscillations because of entropy increase \cite%
{tol}. These oscillating regions will approach flatness asymptotically
unless a strong energy condition violating matter source, like a
potential-dominated scalar field, comes to dominate. In that case the
oscillations will cease and be replaced by unending inflationary expansion 
\cite{BD}. However, studies of the fate of anisotropic closed universes have
shown that the sequence of growing oscillations demanded by the Second Law
of thermodynamics becomes increasingly anisotropic \cite{BG1, BG2}.

The issue of whether a closed region could envelop an open region in
inflationary universes which are still very close to the critical density at
late times was first posed by Zeldovich and Grishchuk \cite{ZN} in the
setting of a spherically symmetric model with $S^{3}$ spatial topology.
This raises the question of the conditions for closed universes to collapse.
It is difficult to answer in general because (unlike, for the singularity
theorems) it involves properties of the general Einstein equations rather
than simply of the geodesic equations. The general fate of closed universes
is addressed by the \emph{closed-universe recollapse conjecture} \cite{BTclosed,
BGT}. It depends upon the spatial topology of the universe. Only spaces
with $S^{3}$ or $S^{2}\times S^{1}$ (and products thereof) can possess
maximal hypersurfaces and hence have an expansion maximum and so collapse.
 These results were recently confirmed in reference \cite{Kleban:2016sqm}.
This is a necessary condition but it is far from sufficient because various
conditions must also apply to the matter content. Surprisingly, closed FRW
universes can avoid recollapse even when $\rho >0$ and $\rho +3P>0$ because
they can experience finite-time infinities in the acceleration of the scale
factor before a maximum is reached, \cite{BGT} (so called `sudden'
singularities \cite{sud}). This can be avoided by imposing a matter
regularity condition, like $\left\vert P\right\vert <C\rho $, with $C>0$
constant or by continuity of $dP/d\rho $. Similar unusual behaviours for
higher time-derivatives of the scale factor are also possible for scalar
fields with fractional power-law potentials \cite{BG}.\footnote{%
A proof for collapse of closed Bianchi type IX universes was given by Lin
and Wald \cite{lin} and other cases with $S^{3}$ and $S^{2}\times S^{1}$
topologies in refs. \cite{BTclosed, BGT}.}

\subsection{Anisotropic spatial curvature}
\label{sec:anisocur}

In our discussion of a possible phase of curvature-dominated
expansion prior to inflation in \S\ref{sec:perts}, we focussed on the simple case of isotropic
negative spatial curvature, which is a familiar ingredient in the Friedman
equation for FRW universes with hyperbolic space sections. However, if we
are interested in the effects of significant levels of inhomogeneity we need
to take into account the effects of anisotropy as well. Cosmological
anisotropies can be due to simple anisotropies in the expansion rate, with
no anisotropy in the spatial curvature, and these are typified by the
Kasner-like behaviour. They contribute an anisotropy energy density, $\sigma
^{2}$, to the Friedman equation that falls off as $a^{-6}$, where $a$ is the
mean expansion scale factor. For these cosmological models the 3-curvature
is of constant sign. For a massless scalar field, $\phi $ in a Kasner metric,
\begin{equation}
ds^{2}=-dt^{2}+t^{2p_{1}}dx^{2}+t^{2p_{2}}dy^{2}+t^{2p_{3}}dz^{2},
\label{kas1}
\end{equation}%
we have
\begin{equation}
\phi (t)=\phi _{0}+\sqrt{\frac{2}{3}}\ln (t),~~\rho =\frac{\dot{%
\phi}^{2}}{2}=\frac{f^{2}}{2t^{2}},~~\sum%
\limits_{i=1}^{3}p_{i}=1=\sum\limits_{i=1}^{3}p_{i}^{2}+f^{2},~~%
0\leq f^{2}\leq 2/3 \, .  \label{kas2}
\end{equation}%
When the constant $f^{2}=0$ this is the vacuum Kasner metric; when $f^{2}=2/3
$, it becomes the isotropic FRW universe \cite{JDBNat}.\footnote{%
The ranges of the $p_{i}$ are not disjoint: $-1/3\leq p_{1}\leq 1/3,0\leq
p_{2}\leq 2/3,1/3\leq p_{3}\leq 1.$The metric has two free parameters, $f$
and one of the $p_{i}.$}

We will show in Appendix \ref{app:anisocur}
that a period of evolution dominated by anisotropic 3-curvature
produce a mean scale factor evolution $a\propto t^{1/(1+2\Sigma )}$ where $%
\Sigma =\sigma /\Theta $ 
is the constant shear ($\sigma$) to volume
Hubble rate ($\Theta$). Notice that the anisotropy falls very slowly ($\sigma \propto
1/t$), compared to the situation in Kasner models with isotropic 3-curvature.
When $\Sigma =0$ we reduce to the isotropic curvature dominated analysis of
pre-inflation in \S\ref{sec:perts}. As the 3-curvature anisotropy increases and $%
\Sigma \rightarrow 1$ the average expansion dynamics that control the onset
of inflation mimic a perfect fluid universe with an energy density 
\[
p/\rho \equiv w=\frac{(4\Sigma -1)}{3}
\]%
with $-1/3\leq w\leq 1/3.$ The case with non-zero anisotropic curvature,
anywhere in the allowed range of values ($0<\Sigma \leq 1$) corresponds to
the effect of a fluid with an equation of state running between that of the
massless scalar field itself ($\Sigma =1$) and that of the isotropic
curvature. This reduces it to the analysis of fluid dominated pre-inflation 
in \S\ref{sec:perts}. 

\section{The severity of the problem of inhomogeneous initial data}

\label{sec:results} In this section, we translate the numerical stability
bound (equation \eqref{eq:bound1}) into a condition on the pre-inflationary
cosmology and the inflationary model. In \S\ref{sec:bound}, we show
that this leads to a novel lower bound on the tensor-to-scalar ratio, $r,$
in decelerating cosmologies with inhomogeneities imprinted when $\Hic=1$. In
\S\ref{sec:hic} we assess the severity of the problem for $\Hic\ll 1$,
and in \S\ref{sec:K} we consider the special case of pre-inflationary
universes dominated by negative curvature.

\subsection{The stability condition}

\label{sec:bound} Inflation fails if any mode is large enough to trigger
destabilisation, so that the probability of a successful period of inflation
is given by, 
\be
{\mathcal P}({\rm inflationary~success}) =  \prod_{|{\bf k}|\geq \hi} {\mathcal P}(|\delta\phi_{\bf k}| < \delta \phi_{\rm max}(k) ) \, ,
\ee 
 where $\delta \phi _{\mathrm{max}}$
denotes the minimum amplitude of inhomogeneities that spoil inflation, and
where ${\bf k}$ runs over the discrete set of Fourier modes within the horizon. In
\S\ref{sec:danger} we reviewed why modes with $k<\hi$ can be neglected, and
also why modes with $k\approx \hi$ are expected to more dangerous for
inflation than those with shorter wavelengths. For concreteness, we take as
our definition of a `stable' model to be one in which, with 95\%
probability, the 
total amplitude of 
field fluctuations in the range $\mathcal{D}_{k}\equiv \lbrack %
\hi,\sqrt{3}\hi]$ do not disrupt inflation according to the condition \eqref{eq:bound1}: 
\be
{\mathcal P}(\delta\phi_{\mathcal{D}_k} < f \Delta \phi ) \geq 0.95 \, .
\label{eq:stabcond}
\ee
The upper limit of ${\cal D}_k$ is chosen to match the `$N=1$' simulations of reference \cite{Clough:2016ymm}.
Equation \eqref{eq:stabcond} neglects the possible failures
due to higher-$k$ modes, but as long as these are rarer than the failure due
to the $k=\hi$ modes, this will result in a small correction to the overall
survival probability for inflation. Moreover, equation \eqref{eq:stabcond}
is the probability of inflationary success of each Hubble-sized patch at $%
\ti
$, but if the pre-inflationary universe involves a large number of such
patches (as is expected in flat and open cosmologies), inflation can succeed
globally despite a low probability of success of each patch. This makes the
condition \eqref{eq:stabcond} a conservative one.

In \S\ref{sec:nic} we parametrised the inhomogeneities at $\tic$ as
Gaussian fluctuations, consistent with the assumptions of references \cite%
{East:2015ggf, Clough:2016ymm}. In the continuous-$k$ approximation, the
variance of the dangerous modes with $k\in \mathcal{D}_{k}$ is given by, 
\be
\sigma^2(\ti)\Big|_{{\cal D}_k} = \int_{{\cal D}_k} \dd \ln k\,  P_{\delta \phi(\ti)}(k) \, .
\ee

We now restrict our discussion to pre-inflationary cosmologies with
decelerated expansion. The power spectrum of field inhomogeneities at $\ti$
is given by, 
\be
P_{\delta \phi(\ti)}(k) = P_{\delta \phi(\tic)}(k) \left( \frac{\hi}{k}\right)^q 
 \, ,
 \label{eq:Pti}
\ee
where the last factor captures the subhorizon evolution of the
inhomogeneities. In the perturbative regime, $q=\tfrac{1}{2}$ for
kinetic-energy domination and $q=\tfrac{2}{1+3w}$ for fluid domination,
cf.~equations \eqref{eq:deltaphikin} and \eqref{eq:deltaphifluid}. In this
section, we focus on a narrow range of modes with $k\approx \hi$ and neglect
this additional damping.\footnote{%
As we will discuss in \S\ref{sec:K}, in the case of negative isotropic curvature-dominated pre-inflationary
cosmologies, this subhorizon damping will become important.} Moreover, in equation \eqref{eq:Pti}, we have assumed that the scalar field inhomogeneities do not evolve substantially on superhorizon scales (consistent with our findings in \S\ref{sec:preinf}). 
 The variance is then given by,
\be
\sigma^2(\tau_i) \Big|_{{\cal D}_k} = A\, b \left( \frac{\hi}{\hic}\right)^{\nic-1} 
=
A\, b \left( \frac{\Hi\, a_{\rm i}}{
H_{\rm IC}\, a_{\rm IC}
}\right)^{\nic-1} 
\, , 
\label{eq:Q2}
\ee
where we have defined,
\be
b  = \frac{\sqrt{3}^{\nic-1}-1}{\nic -1} \, .
\label{eq:b}
\ee
As $\nic\to1$, $b\to \ln\sqrt{3}$; unless the spectrum is very steep, $b$ tends not to be very large. The amplitude  $A$ is defined in equation \eqref{eq:nic}, and we recall that $A=1$ (which we will use as a `generic' value in our estimates below) corresponds to $\delta \phi \sim {\cal O}(1)$ for modes with $k=\hic$.

Equation \eqref{eq:Q2} depends on the energy scale of inflation and the
amount of expansion between $\tic$ and $\ti$. To make analytic progress, we 
parametrise the decreasing energy density by a single-fluid
equation-of-state parameter $w$, which we take to be constant during the
pre-inflationary epoch. We note that this parametrisation includes, but is
not limited to, the examples of decelerating cosmologies discussed in
\S\ref{sec:perts}. 
At $\ti$, the scalar potential has just begun to dominate the energy density
so that,\footnote{%
In highly inhomogeneous cosmologies, the energy density and Hubble parameter
can be defined through spatial averaging, cf.~\cite{East:2015ggf,
Clough:2016ymm}.} 
\be
\frac{a_{\rm i}}{a_{\rm IC}} = 
\left( 
\frac{\rho_{\rm IC}}{\rho_{\rm i}}
\right)^{\frac{1}{3(1+w)}} 
=
\left( 
\frac{H_{\rm IC}}{H_{\rm i}}
\right)^{\frac{2}{3(1+w)}} 
\, .
\ee
The variance of the scalar field inhomogeneities then simplifies to,
\be
\sigma^2(\tau_i) \Big|_{{\cal D}_k} 
=
A\, b \left( \frac{\Hi}{
H_{\rm IC}
}\right)^{p} 
\, ,
\label{eq:Q3} 
\ee
where,
\be
p = (\nic -1) \frac{1+3w}{3(1+w)} \, .
\ee 

Assuming that the amplitude $\delta \phi^2_{\mathcal{D}_k}$ is the square of a Gaussian with variance $\sigma^2 =\sigma^2 (\ti)|_{\mathcal{D}_{k}}$,
%
then with $95\%$ probability, $|\delta \phi _{\hi%
}|<2.0\sigma $.\footnote{If $n$ independent Gaussian modes contribute to the amplitude, then $\delta \phi^2_{\mathcal{D}_k} \sim \sigma^2 \chi^2_n$ so that $\sigma^2 = \tfrac{1}{n}\sigma^2 (\ti)|_{\mathcal{D}_{k}}$. This leads to marginally stronger bounds that we will not consider here. } Combined with the stability condition %
\eqref{eq:stabcond}, this  gives, 
\be
2.0\, \sigma(\tau_i) \big|_{{\cal D}_k} < f \phii \, .
\label{eq:bound2}
\ee
We take equation \eqref{eq:bound2} as our
condition for when a model is said to have no problem with initial inhomogeneities.
Using equations \eqref{eq:scaling} and \eqref{eq:Q3}, this leads to our main
inequality: 
\be
\left( \frac{H_{\rm IC}  }{H_{\rm i} }\right)^{2-p}  <  \frac{f^2 \phiref^2}{4A\, b} \left(\frac{H_{\rm IC}}{\Href}\right)^2 \, .
\label{eq:bound}
\ee

Equation \eqref{eq:bound} captures the competition between the two relevant
effects: the lower the energy scale of inflation, the 
larger is the ratio $\hic/\hi$, but also, 
 the narrower is the inflationary plateau. This competition leads
to two qualitatively different regimes depending on whether the power
spectra are `steep' with $p\geq 2$, or `moderate' with $0<p<2$. We now
discuss the implication of equation \eqref{eq:bound} for each of these
possible cases.

\subsection{A lower bound on $r$}

For spectra with $p>2$, equation \eqref{eq:bound} implies an \emph{upper}
bound on the inflationary energy scale, 
\bea
\log \Hi 
<  \frac{1}{p-2} \log\left(
\frac{f^2 \phiref^2}{4 Ab \Href^2} \Hic^p
\right) \, .
\eea
 Taking $\phiref$ and $\Href$ as in equation \eqref{eq:ref}, setting $f=1.6$ as discussed around equation \eqref{eq:cond1},
and taking 
$A=\Hic=1$,  and, for concreteness, $b=1$
(neglecting its parameter dependence), we find, 
\be
\log \Hi <   \frac{1}{p-2} \log \left(6.2\times10^{14} \right) \, .
\ee
 Since
the right-hand-side is non-negative (for $p>2$), but $\log \Hi<0$ in any
inflationary model, this inequality does not constrain $\Hi$. 
For $p=2$, the dependence on the inflationary energy immediately cancels
between the two competing effects, and $\Hi$ is again unconstrained. In
other words, regardless of the energy scale of inflation, there is no
problem with inhomogeneous initial data if their spectrum is steep and
imprinted at the Planck scale. 

 \begin{figure}
 \centering
 \includegraphics[width=.85\textwidth]{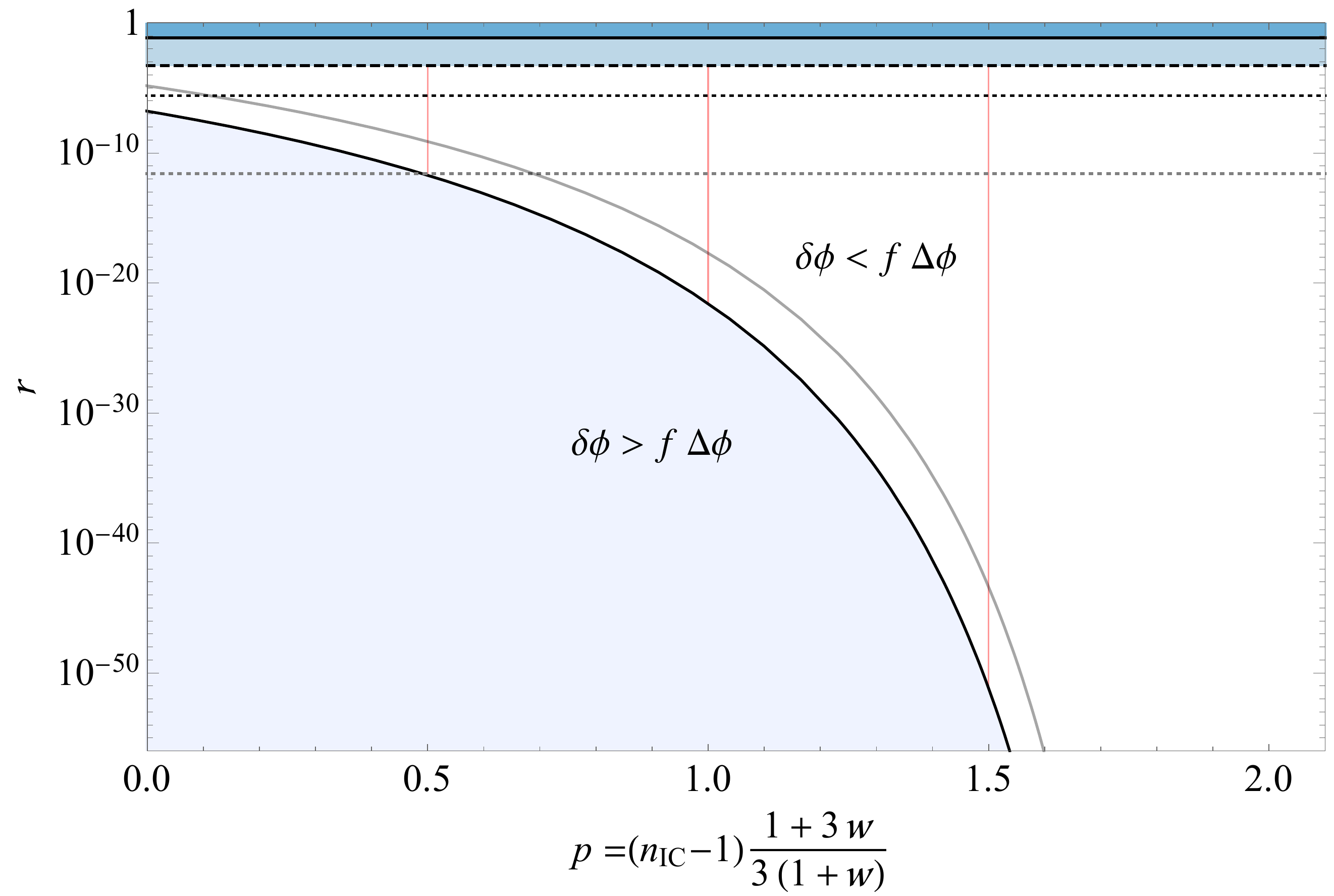}~~
 \caption{A lower bound on $r$: models with $\delta \phi < f \Delta \phi$
 do not have a problem with inhomogeneous initial  data imprinted at the Planck scale. The black curve corresponds to inflection-point potentials with a flat extension to negative field values ($f=1.6$); the grey curve corresponds to sharply rising  potentials ($f=0.17$). 
Dotted horizontal lines, from top down, correspond to the boundary of the small-field region and the `typical' model of reference \cite{Clough:2016ymm}.
Shaded  regions on top correspond to current observational constraint from BICEP2/Planck, $r<0.07$ (dark blue) and possible future constraint from CMB Stage-4 experiment with $\sigma(r)\approx 5\times 10^{-4}$ (lighter blue). 
 Vertical red lines correspond to $w=1/3$ and, from left to right,  $\nic =2,\, 3,\, 4$. 
  }
  \label{fig:boundr}
 \end{figure}

For spectra with a `moderate' fall-off, equation \eqref{eq:bound} implies an interesting  \emph{lower} limit on $\Hi$:
\be
\log \Hi > -\frac{1}{2-p} \log\left( 
\frac{f^2 \phiref^2}{4 Ab \Href^2} \Hic^p
\right) \, .
\ee
We may express this as a bound on $r$ by using,
\be
\log_{10} r = \log_{10} P_t - \log_{10} P_s =
8.0 + 2 \log_{10} \Hi \, ,
\ee
where  we have used the normalisation of the primordial power spectrum, $P_s=2.2\times10^{-9}$, and $P_t = \tfrac{2}{\pi^2} \Hi^2$.
This gives,
\be
\log_{10} \ri > 8.0 - \frac{2}{2-p}\log_{10}\left(\frac{f^2}{4 A b}\frac{\phiref^2}{\Href^2}H_{\rm IC}^p \right) \, .
\label{eq:boundr}
\ee 
Again taking $\phiref$ and $\Href$ as in equation \eqref{eq:ref} and setting $A= \Hic=b=1$ and $f=1.6$,  we find the limit, 
\be
\log_{10} \ri > 8.0 - \frac{29.6}{2-p} \, .
\label{eq:rexpl}
\ee
Equation \eqref{eq:rexpl} is plotted in Figure \ref{fig:boundr}.
Models with tensor-to-scalar ratios above  this limit have no problem with inhomogeneous initial  data.
We note that this includes rather vast regions of parameter space. For example,  
equation \eqref{eq:rexpl} implies that initial inhomogeneities with $\nic=3$ that pass through a radiation dominated pre-inflationary era from the Planck scale are not problematic for inflationary models with,
\be
\ri \big|_{p=1} > 2.5\times 10^{-22} \, .
\ee
For potentials raising steeply for negative values, $f=0.17$ as discussed following equation \eqref{eq:cond1}, and the constraint \eqref{eq:rexpl} becomes more severe: $\log_{10} \ri > 8.0 - \frac{25.7}{2-p}$ so that $r|_{p=1} >2.0 \times 10^{-18}$.  

 The weakness of these lower bounds
indicates that the inhomogeneous initial data problem is not in general
severe even for small-field inflection-point models. It should be noted that
the current Planck constraint (or even a hypothetical stronger, future
constraint from ground-based CMB experiments) does not significantly impact
the inhomogeneous initial data problem, contrary to the assertions of \cite%
{Ijjas:2013vea}.

For our `typical' reference model of equation \eqref{eq:ref}, the bound \eqref{eq:bound} can be written as a constraint on $p$:
\be
p > \frac{\log\left( \frac{f^2 \phiref^2}{4 A b}\right)}{\log \Href}  =0.49 \, ,
\ee
where in the last step we have specialised to $A=b=1$ and $f= 1.6$.
Consequently, for a radiation dominated pre-inflationary era, the reference model is safe from $\delta \phi_{k=\hic}(\tic)\sim 1$ inhomogeneities if $\nic\geq2.0$.

\subsection{Initial data imprinted below the Planck scale}

\label{sec:hic} The energy density at the initial time $\tic$ is bounded
from above by the Planck density, but $\Hi<\Hic\ll 1$ is a general
possibility. If the initial inhomogeneities were imprinted at energies much
below the Planck scale, the pre-inflationary phase is shortened, and the
initial data problem can become more severe.

\begin{figure}
 \centering
 \includegraphics[width=.85\textwidth]{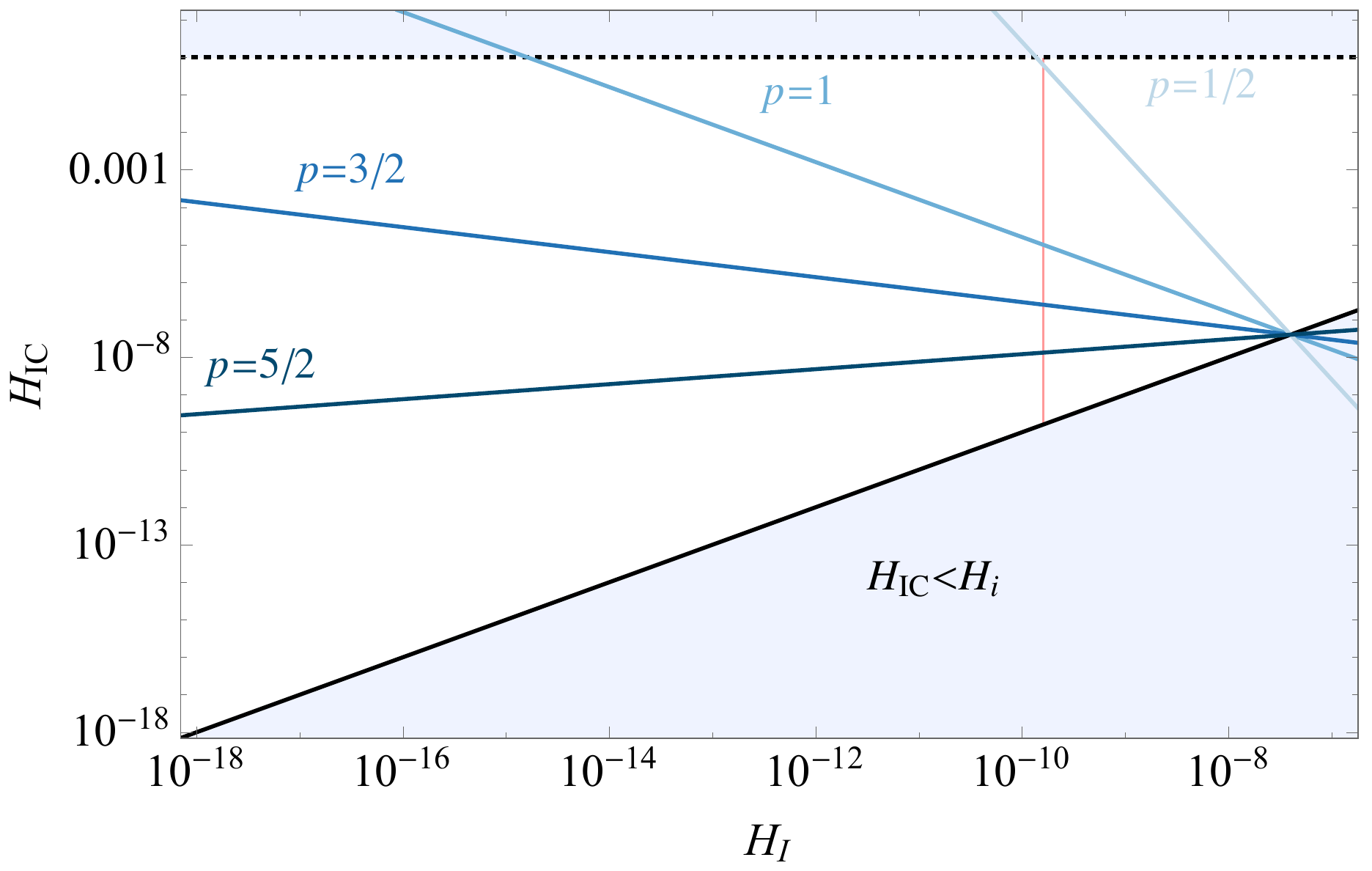}~~
 \caption{The bounds on  $\Hic$ above which initial inhomogeneities with unit amplitude fail to disrupt inflation.The vertical red line corresponds to the reference model. 
  }
  \label{fig:HIC}
 \end{figure}

Using equation \eqref{eq:bound}, we see that avoiding the inhomogeneous
initial data problem requires an initial Hubble rate satisfying, 
\be
H_{\rm IC} >  \left( \frac{4 Ab\Href^2}{f^2 \phiref^2}\right)^{1/p} \frac{1}{\Hi^{(2-p)/p}} \, ,
\ee 
 Figure \ref{fig:HIC}
shows the bound on $\Hic$ for the parameters $A=b=1$, $f=1.6$ and $\phiref$
and $\Href$ as in equation \eqref{eq:ref}. 
For moderate spectra with $p<2$, we find that for $\Hic<4.0\times 10^{-8}$,
the entire family of small-field models are susceptible to disruption from
inhomogeneities. Models with steep initial spectra ($p>2$) are more robust.
For our reference small-field model and a pre-inflationary epoch with $p=1$
(e.g.~given by a radiation domination and inhomogeneities with $\nic=3$),
the initial data problem is absent as long as $\Hic>1.0\times 10^{-5}$.

\subsection{Isotropic negative-curvature domination}

\label{sec:K} In the case that the pre-inflationary universe is dominated by
isotropic negative curvature, we have that $\epsilonH=1$ and the comoving
Hubble radius is constant. The `dangerous' modes are then within the horizon
during the entire pre-inflationary epoch. Using the perturbative damping of
equation \eqref{eq:Kdamping}, and taking $\sigma ^{2}(\tic)\big|_{\mathcal{D}%
_{k}}=Ab$ with $b$ as in equation \eqref{eq:b}, equation \eqref{eq:bound2}
gives,
 \be
 \Hic > \frac{2  \sqrt{Ab}}{f \phiref}\, \Href \, .
  \ee
 This
inequality is the same as that obtained for a decelerating pre-inflationary
epoch with $p=2$, and is clearly independent of $\Hi$. Taking $A=b=1$, $f=1.6$,
and using \eqref{eq:ref}, we again find the bound, 
\be
 \Hic > 4.0\times10^{-8} \, .
 \ee
 Initial inhomogeneities imprinted above this energy scale
are not expected to disrupt inflation.

\section{Conclusions}

\label{sec:conclusons} \label{sec:concl} We have studied the inhomogeneous
initial data problem in small-field inflection-point models for inflation.
These models only require the potential to be flat over a very limited field
range $\Delta \phi \ll 1$, and are sensitive to disruption from scalar-field
inhomogeneities with an amplitude $\delta \phi \sim \mathcal{O}(\Delta \phi
) $ \cite{East:2015ggf, Clough:2016ymm, Clough:2017efm}. The problem of
inhomogeneities is expected to be worse for this class of models than those
with extended field ranges. 

We have emphasised that the time at which the initial data were imprinted, $%
\tic$, may have greatly preceded the time $\ti$ of the would-be onset of
inflation, and that the most dangerous modes for disrupting inflation had
wavelengths far bigger than the horizon at $\tic$. By parametrising the
initial inhomogeneities at $\tic$, and the pre-inflationary evolution, we
have used the numerically derived condition $\delta \phi \big|_{k\approx \hi%
}\leq f\Delta \phi $ to find those cosmological models that free from the
disruptive inhomogeneous initial data problem. The resulting simple bound
depends only on a few parameters of the pre-inflationary cosmology, and the
energy scale of inflation.

For inhomogeneities with $\delta \phi (\tic)\big|_{k=\hic}\sim 1$ imprinted
at the Planck scale, we expressed this bound as a lower limit on the
tensor-to-scalar ratio, $r$. For many pre-inflationary cosmologies, this
limit is very weak, which indicates that inflection-point inflation do not 
\emph{in general} possess a problem with inhomogeneous initial data. In
particular, these results give no credence to the envisioned doom of the
inflationary scenario, proposed in reference \cite{Ijjas:2013vea}, through
the unnaturalness of the inhomogeneous initial data problem for small-field
inflation. 

Unless the initial spectrum of inhomogeneities falls off very steeply for
super-horizon wavelengths and the pre-inflationary expansion is
decelerating, the initial data problem becomes more severe if $\Hic$ is much
below the Planck scale. We find that all models in this class fail for $\Hic%
\lesssim 4\times 10^{-8}$.

Our analysis has a number of caveats. 
We have focussed on distributions of inhomogeneities that
decrease in amplitude on scales far larger than the horizon (i.e.~$\nic>1$),
which is physically well-motivated as we discussed in \S\ref{sec:nic}.
For $\nic\leq 1$ (over some large but finite range of scales, so that the total power stays finite), the inhomogeneous initial data problem is expected to be
more severe. Such a scenario is conceivable if the primary source of the initial inhomogeneities is much larger the initial Hubble radius.

Moreover, in \S\ref{sec:perts} we reviewed the sub-horizon decay of
scalar field inhomogeneities for small perturbations around a homogeneous
background cosmology, and  in Appendix \ref{sec:Liang}  we reviewed how similar results hold for the special case of nonlinear
cylindrically symmetric scalar field inhomogeneities. For decelerating pre-inflationary backgrounds, modes that just
enter the horizon at the onset of inflation are most dangerous, and the sub-horizon
fall-off rate of small-wavelength inhomogeneities is not very important.
However, for isotropic negative curvature dominated cosmologies, the
dangerous modes are always inside the horizon. We expect that nonlinear evolution involving
caustics, shock waves and gravitational collapse  leads to a
more rapid decay of sub-horizon inhomogeneities, thereby making
inflation more robust than our estimates suggest.

Our estimates neglect the evolution of the scalar field inhomogeneities on super-horizon scales. Since this is a very small effect in both the perturbative cosmologies and the non-linear, cylindrically symmetric case, we expect our results to be rather broadly applicable. However, pre-inflationary cosmologies  in which $\delta \phi_k$ is boosted on superhorizon scales can lead to more stringent bounds than those we derive. 

Future numerical work could improve our bounds. The chance that modes with $%
k\ll\hi$ destabilises inflection-point inflation was only briefly studied in 
\cite{Clough:2016ymm} (see also \cite{East:2015ggf}). A determination of how
the coefficient $f$ in equation \eqref{eq:bound1} grows with $k$ could
directly improve the determination of the full destabilisation probability.
Moreover, references \cite{East:2015ggf} and \cite{Clough:2016ymm} studied
the impact of modes with $k\geq\hic$ on inflation. A full simulation of the
inhomogeneous pre-inflationary phase preceding inflection-point inflation,
including modes with $\hi \leq k \leq \hic$, would provide an independent
test of the bounds derived in this paper. Finally, our work has focused
exclusively on single-field inflation, and it is possible that multi-field
models of small-field inflation are more sensitive to initial
inhomogeneities, leading to interesting constraints on the combination of
the number of fields and the tensor-to-scalar ratio. Recent advances in
constructing explicit inflationary models with many interacting fields \cite%
{Marsh:2013qca,Dias:2016slx,Dias:2017gva,Bjorkmo:2017nzd} could allow for
this question to be investigated in detail.

\subsection*{Acknowledgements}
We would like to thank Eugene Lim, Josu Aurrekoetxea, Katy Clough, and Raphael Flauger for comments on an draft of this paper. 
J.D.B. is supported
by the Science and Technology Facilities Council (STFC) of the UK. C.G. is
supported by the Jawaharlal Nehru Memorial Trust Cambridge International
Scholarship.
D.M. is supported by a Stephen Hawking Advanced Fellowship at the Centre for
Theoretical Cosmology, DAMTP, University of Cambridge. 

\begin{appendix}

\section{Perturbative pre-inflationary inhomogeneities}
\label{app:pert}
In this appendix, we provide details of the derivation of the general evolution of scalar field inhomogeneities summarised by equation \eqref{eq:decayrate}. We consider in turn each of the cases listed in equation \eqref{eq:classes}. 

\subsubsection*{\textit{i)} Kinetic-energy domination}

In the limit where the scalar field's kinetic energy dominates the total
energy density (and $\mathcal{K}$ is negligible), the equations for the
metric perturbations, cf.~\eqref{eq:perts1}--\eqref{eq:perts3}, simplify to, %
\bea
\frac{1}{{\cal H}^2} \nabla^2 \Phi - 3  \Phi' =
 \frac{ \sqrt{6}}{2} \delta \phi'  \, , ~~
  \Phi'' +   \Phi' = \frac{\sqrt{6}}{2}\delta \phi'  \, ,
  \label{eq:KEpert}
 \eea
 where $X^{\prime }=\dd X/\dd N\equiv %
\dd X/\dd(\mathcal{H}\tau )$ denotes a derivative with respect to the number
of e-folds. 
Re-expressed in flat-space Fourier modes, 
\be
\Phi_{\bf k}(\tau) = \frac{1}{(2\pi)^{3/2}} \int \dd^3 x \Phi(\tau, x)\, e^{-i k \cdot x} \, ,
\ee
the  solutions of equation \eqref{eq:KEpert} are given by the Bessel functions,
 \be
 \Phi_{\bf k}(\tau) = {\cal H}(\tau) \left(c_1\,  J_1\left(\frac{k}{2 {\cal H}}\right) 
 + c_2\,  Y_1\left(\frac{k}{2 {\cal H}}
 \right) \right)
 \, .
 \ee
For small arguments, $\tfrac{k}{2\mathcal{H}}\ll 1$, $\Phi _{\mathbf{k}%
}(\tau )$ has a constant and a decaying solution. 
Dropping the latter, we note that both the Newtonian potential and the
scalar field perturbation are constant for $\tfrac{k}{2\mathcal{H}}\ll 1$. 

Upon `horizon entry' at $k=2\h$, the modes begin damped oscillations. In the
large argument expansion, $J_{\nu }(x),~Y_{\nu }(x)\sim 1/\sqrt{x}\times
\cos \left( x-\vartheta \right) $, with $\vartheta $ depending on $\nu $ and the
type of Bessel function. It then follows that the envelopes of the
gravitational potential and the scalar field decay as, 
\be
\Phi_{\bf k} \sim \left( \frac{{\cal H}}{k}\right)^{3/2} \, ,~~\delta \phi_{\bf k} \sim e^{-N} \sim  \left( \frac{{\cal H}}{k}\right)^{1/2} \, .
\label{eq:deltaphikin}
\ee

\subsubsection*{\textit{ii)} Fluid domination}
When the fluid dominates the energy density (and $\mathcal{K}$ is
negligible), the gravitational potential $\Phi $ is governed by the equation, 

 \bea
  \Phi_{\tau \tau} + 3(1+ w) {\cal H} \Phi_{\tau} - w \nabla^2\Phi  &=& 0 \, . \label{eq:pertfluid}
 \eea
 Since we are considering a perfect, isentropic fluid,  $\delta P = w \delta \rho$, and the sound-speed is given by $w = c_s^2$.
%
%
 %
 %
Expressed in terms of the Fourier modes, $\Phi_k(\tau)$, the  solutions to equation \eqref{eq:pertfluid}  are the Bessel functions,
 \be
 \Phi_k(\tau ) = \h^{\nu} \left(
c_1\, J_{\nu}\left( \frac{2\sqrt{w}}{1+3 w}  \frac{k}{{\cal H}}\right)
 +c_2\, Y_{\nu}\left( \frac{2\sqrt{w}}{1+3 w}   \frac{k}{{\cal H}}\right)
 \right) \, ,
 \ee
 for $\nu = \tfrac{1}{2} (5+3w)/(1+3w)$. In the long wave-length limit,
 $
 k/\h \ll  \frac{2\sqrt{w}}{1+3 w}
 $, 
these  solutions are again given by one constant and one decaying mode. 
By contrast,  short wave-length `sound waves' oscillate and decrease in magnitude as,
\be
\Phi_k(\tau) \sim  \left(\frac{\h}{k}\right)^{\nu + \frac{1}{2}} \, .
\ee
The exponent is minimised as $w  \to 1$, in which case the scaling agrees with the kinetic energy dominated case: $\Phi_{\bf k} \sim \left({\cal H}/k\right)^{3/2}$. For $w=1/3$, $\Phi_k \sim ({\cal H}/k)^{2}$. 
 The scalar field perturbations are governed by the Klein-Gordon equation, 
which in this limit  is given by,
 \be
\delta \phi_{\tau \tau} + 2 \h \delta \phi_{\tau} - \nabla^2 \delta\phi = 0 \, .
\ee
The solutions for the Fourier modes can be expressed in terms of the Bessel functions,
\be
\delta \phi_{\bf k} = 
\h^{-\tilde \nu} \left( c_1 J_{\tilde \nu} \left( \frac{2}{1+3w} \frac{k}{\h}\right)
+  c_2 Y_{\tilde \nu} \left( \frac{2}{1+3w} \frac{k}{\h}\right)
\right) \, ,
\ee
where $\tilde \nu =\tfrac{3}{2}\tfrac{w-1}{1+3w}<0$. 
Again, for $k^2 \ll {\cal H}^2$, there exists a constant solution for $\delta \phi_k(\tau)$. For $k\gg {\cal H}$, $\delta \phi_{\bf k}$ goes through damped oscillations and decays like,
\be
\delta \phi_{\bf k} \sim e^{-N} \sim \left(\frac{\h}{k} \right)^{\frac{2}{1+3w}} \, .
\label{eq:deltaphifluid}
\ee 


\subsubsection*{\textit{iii)} Isotropic curvature domination}
In the negative-curvature dominated universe, the comoving Hubble parameter
is constant and given by, 
 \be
{\cal H}^2 = -{\cal K} \, .
\ee
The gravitational scalar perturbations are governed by the equation, 
\be
 \Phi'' 
 + 6  \Phi'
  -
 \frac{1}{{\cal H}^2} \nabla^2 \Phi 
  + 8  \Phi = 0\, .
 \label{eq:Kperts}
\ee
Since the
three-dimensional hypersurfaces are open and negatively curved, we cannot
expand the solution in plane waves. Following \cite{Lifshitz:1963ps,
Wilson83}, we instead use the directly analogous expansion of $\Phi $ in
terms of eigenfunctions of the Laplace operator on negatively curved spaces,
which are characterised by their wave-number $k$ satisfying $\nabla ^{2}\Phi
_{k}=-k^{2}\Phi _{k}$ for $k^{2}\geq -K$. Expressed in terms of this basis
of functions, equation \eqref{eq:Kperts} becomes, 
\be
 \Phi''_k 
 + 6  \Phi'_k 
  +\left( 8+
 \frac{k^2}{{\cal H}^2} \right) \Phi_k 
  = 0\, .
\ee
Since ${\cal H}$ is constant during this era, the solutions are simply given by,
\be
\Phi_k = e^{-3N} \left( 
c_1 e^{-i N \sqrt{\tilde k^2-1}} + c_2  e^{i N \sqrt{\tilde k^2-1}} 
\right) \, ,
\label{eq:Ksol}
\ee 
where $\tilde k^2 = k^2/{\cal H}^2 \geq 1$.

The scalar field perturbations are determined by equation \eqref{eq:perts2}, 
\be
\frac{1}{2} \bar \phi' 
\delta \phi_{\bf k} = \Phi_k + \Phi'_k \, .
\ee
Since $\bar{\phi}^{\prime }\sim \exp (-2N)$
(cf.~equation \eqref{eq:bkgspeed} for $w=-1/3$), the scalar field
perturbation oscillates with an amplitude that decreases like, 
\be
\delta \phi_{\bf k} \sim e^{-N} \sim \frac{H}{\Hic}\, ,
\label{eq:Kdamping}
\ee
 for all $k$.

Therefore, in all of the pre-inflationary cosmologies that we consider, the
perturbative scalar field inhomogeneities behave like, 
\bea
\delta \phi_{\bf k} \sim \left\{
\begin{array}{l c l}
{\rm constant} &~~~& k\ll {\cal H}\, , \\
e^{-N} &~~~ & k \gg {\cal H} \, .
\end{array}
\right.
\label{eq:deltaphi}
\eea
In pre-inflationary cosmologies with
successive eras during which the cosmic energy density is dominated by
different sources, the super-horizon modes of $\Phi _{\mathbf{k}}$ and $
\delta \phi _{\mathbf{k}}$ evolve only mildly during transition periods (in
direct analogy to the shift in superhorizon modes of $\Phi $ by $9/10$ at
matter-radiation equality after inflation).

Large inhomogeneities are not captured by this analysis. 
In some special cases with large and highly symmetric inhomogeneities, the perturbative results extend straightforwardly, as we  exemplify in Appendix \ref{sec:Liang}. 
More general non-linear inhomogeneities may undergo
prompt gravitational collapse 
after they enter the horizon, leading to a quicker decay of long-wavelength
inhomogeneities than equation \eqref{eq:deltaphi} suggests.

\section{Non-linear pre-inflationary inhomogeneities}
\label{app:nonlinear}
In this appendix, we bring together relevant results on the evolution of non-linear inhomogeneous cosmologies that admit exact solutions to Einstein's equations. 
In Appendix \ref{sec:Liang}, we show that cylindrical scalar field inhomogeneities evolve much like small perturbations around the FRW solutions,  and in Appendix \ref{app:anisocur} we provide further details on pre-inflationary universes with anisotropic, negative spatial curvatures.  

\subsection{Large, cylindrical inhomogeneities}
\label{sec:Liang}

In this section we examine the fate of a special class of nonlinear inhomogeneities in the
scalar field. When the metric possesses cylindrical symmetry it is possible
to solve the problem exactly, as first discovered for the vacuum problem by
Einstein and Rosen \cite{Einstein:1937qu}. We take the metric to be,
\begin{equation}
ds^{2}=-e^{2(\chi -\psi )}(d\tau^{2}-dr^{2})+a^2(e^{2\psi }dz^{2}+e^{-2\psi
}(r d\theta)^{2}) \, , \label{ER}
\end{equation}%
where the free functions are $a^2(\tau,r),\psi (\tau,r)$ and $\chi (\tau,r).$ We
 ignore $\psi $ as it contains the gravitational-wave behaviour.

If we add
a massless, homogeneous scalar field $\bar \phi (\tau)$, we recover the flat, kinetic-energy dominated FRW universe as the
special case $\psi=0$, $a^2 = {\rm exp}(2\chi) = \tau$ so that,
\begin{equation*}
p=\rho = \frac{1}{2}e^{-2\chi }\bar \phi_{\tau}^{2} 
=\rho(\tic) \left( \frac{\tic}{\tau}\right)^3
 \, .
\end{equation*}
The background value of the field grows like
$\bar\phi \sim \ln (\tau)$, consistent with equation \eqref{eq:bkgspeed}.

The inhomogeneous exact solution for the metric \eqref{ER} generalises this
to give decoupled  Bessel equations for $\phi (r,t)$ and $\psi (r,t)$ The
solution for $\phi (r,t)$ is \cite{Liang}:
\begin{equation}
\phi (\tau,r)=\bar \phi(\tau)+\frac{1}{(2\pi)^{3/2}}\int_{-\infty }^{\infty }{\rm d} k\, e^{ikr}%
\phi_{k}(\tau) \, ,  \label{sol1}
\end{equation}%
where the Fourier modes are given by,
\begin{equation}
\phi _{k}(\tau)=c_{1}J_{0}(k\tau)+c_{2}Y_{0}(k\tau) \, .  \label{sol2}
\end{equation}


Qualitatively, the evolution is straightforward and identical to that of perturbative inhomogeneities. The ratio of the coordinate
size of the inhomogeneity wavelength to the particle horizon coordinate size
is $k\tau$. On large (superhorizon) scales, where $k\tau\rightarrow 0$, we have $%
J_{0}\rightarrow 1$ and $Y_{0}\rightarrow \ln (k\tau)$, so the evolution is
Kasner-like ($\psi \propto \ln (\tau)\propto \phi $) or FRW-like ($\psi =0$)
depending on whether $c_{2}/(2\pi)^{3/2}$ is bigger or less than the first (homogeneous)
term on the right-hand side of equation \eqref{sol1}.
 Superhorizon inhomogeneities imprinted at $\tau_{\rm IC}$ in general include a constant and a decaying  mode, precisely as we found in \S\ref{sec:perts} for perturbative inhomogeneities.

When the inhomogeneity
enters the horizon (which equals the Jeans length) the future evolution is 
given by the small-scale $k\tau\rightarrow \infty $ limit of the Bessel
functions. 
Due to its restrictive symmetries, this
solution does not admit bound regions and gravitational collapse. Moreover, the scalar
field waves propagate at the speed of light so there can be no shocks.
Precisely as in the case of perturbative inhomogeneities, 
the scalar field inhomogeneities are pressure-damped away through
 oscillatory decay:
$
\phi _{k}(\tau)\simeq \frac{1}{\sqrt{k\tau}}\times \sin(k\tau- \vartheta)
$.
Thus, also in this case,
\bea
\delta \phi_{\bf k} \sim \left\{
\begin{array}{l c l}
{\rm constant} &~~~& k\ll {\cal H}\, , \\
e^{-N} &~~~ & k \gg {\cal H} \, .
\end{array}
\right.
\eea
 Further details of the evolution of
this exact solution and the cosmological case with radiation can be found in \cite{Liang}.

\subsection{Anisotropic spatial curvature}
\label{app:anisocur}

In \S\ref{sec:anisocur} we reviewed how anisotropic negative curvature leads to an average expansion rate that mimics that of a perfect fluid with an effective equation of state in the range $-1/3 \leq w\leq 1/3$. In this section, we provide further details of this analysis. 

Apart from anisotropies in the expansion rate (discussed in \S\ref{sec:anisocur}), 
there can also be anisotropies in the
3-curvature. This is the most general form of anisotropy and is
non-Newtonian, deriving from the magnetic part of the Weyl curvature.
Cosmological models with anisotropic 3-curvature can have very complex
evolution, with chaotic dynamics near the initial singularity and the sign
of the 3-curvature scalar can be time-dependent. Here we outline the new
effects created by anisotropic 3-curvature.

The most general anisotropic Bianchi universes that contain the open
Friedmann model as a special subcase are those of type $VII{}_{h}$. The
late-time asymptotes for the non-tilted type $VII{}_{h}$ spacetimes, with $%
h\neq 0$ and a matter content that obeys the strong energy condition (so no
inflation), evolve towards the vacuum plane-wave metric found by
Doroshkevich et al and Lukash~\cite{DLN, Lukash} that is known as the Lukash metric.
These spacetimes describe the most general effects of spatially homogeneous
perturbations on open FRW universes. When the strong energy condition is
obeyed, then isotropic expansion was found to be stable but not
asymptotically stable at late times~~\cite{CH,B1,BTclosed}.

The line element of the Lukash metric takes the form 
\begin{equation}
{\rm d}s^{2}=-{\rm d}t^{2}+t^{2}{\rm d}x^{2}+t^{2r}{\rm e}^{2rx}\left[ (A%
{\rm d}y+B{\rm d}z)^{2}+(C{\rm d}y+A{\rm d}z)^{2}\right] \,,  \label{DLN}
\end{equation}%
where $r$ is an arbitrary constant parameter in the range $0<r<1$, $A=\cos v$%
, $B=f^{-1}\sin v$, $C=-f\sin v$ and $v=k(x+\ln t)$~\cite{BS2,HW,HSUW}. Note
that $f$, $k$ and $r$ are related by 
\begin{equation}
\frac{k^{2}}{f^{2}}(1-f^{2})^{2}=4r(1-r)\hspace{15mm}{\rm and}\hspace{15mm}%
r^{2}=hk^{2}\,,  \label{Lcons}
\end{equation}%
where $h$ is the associated group parameter. For $r=1$ and $f^{2}=1$ the
Lukash metric reduces to the empty isotropic Milne universe, with scale
factor $a=t$. More details can be found in ref. \cite{btsag} and other
universes with anisotropic curvature can be seen, along with their effects
on primordial nucleosynthesis in \cite{aniso}.\footnote{ If the spatial topology of type $VII_h$, or even type $V$, open universes is made compact then they are all constrained to be isotropic \cite{2001CQGra..18.1753B}.}

If we use the average scale factor ($a$) to define the volume expansion rate
via the standard relation $\Theta =3\dot{a}/a$, then the average volume
expansion of the vacuum Lukash universe is described by the following
version of the Raychaudhuri equation 
\begin{equation}
\dot{\Theta}=-{\textstyle{\frac{1}{3}}}\Theta ^{2}-2\sigma ^{2}\,,
\label{Ray1}
\end{equation}%
where $\sigma ^{2}=\sigma _{ab}\sigma ^{ab}/2$ is the magnitude of the shear
tensor. 

The absence of matter means that the Lukash spacetime is Ricci flat. The
curvature of the spatial sections, however, is not zero. In particular, the
3-Ricci tensor (${\cal R}_{ab}$) is completely determined by its scalar and
its symmetric and trace-free parts, that is respectively by 
\begin{eqnarray}
{\cal R} &=&-{\textstyle{\frac{2}{3}}}\Theta ^{2}+2\sigma ^{2}\,,
\label{LcR} \\
{\cal S}_{ab} &=&-{\textstyle{\frac{1}{3}}}\Theta \sigma _{ab}+\sigma
_{c\langle a}\sigma ^{c}{}_{b\rangle }+E_{ab}\,,  \label{LcSab}
\end{eqnarray}%
where ${\cal S}_{ab}={\cal R}_{\langle ab\rangle }={\cal R}_{(ab)}-{\cal R}%
h_{ab}/3$.\footnote{%
Angled brackets denote the symmetric and trace-free part of orthogonally
projected tensors and the orthogonally projected components of vectors.} The
scalar ${\cal R}$ is negative, which means that the model is spatially open.
The expression (\ref{LcR}) is the generalised Friedmann equation.

The magnitude of the shear tensor associated with the Lukash solution is 
\begin{equation}
\sigma ^{2}={\textstyle{\frac{1}{2}}}\sigma _{\alpha \beta }\sigma ^{\alpha
\beta }=\frac{(1-r)(1+2r)}{3t^{2}}\, .  \label{Lsigma2}
\end{equation}
The mean Hubble volume expansion of the Lukash universe is determined by the
scalar 
\begin{equation}
\Theta =\frac{1+2r}{t}\,.  \label{LTheta}
\end{equation}%
Hence, the average scale factor obeys the simple power law $a\propto
t^{(1+2r)/3}$ with $0<r<1$.  

When measuring the average anisotropy of the expansion, it helps to
introduce the following dimensionless and expansion-normalised shear
parameter 
\begin{equation}
\Sigma \equiv \frac{3\sigma ^{2}}{\Theta ^{2}}\,.  \label{Sigma}
\end{equation}%
In the Lukash spacetime the scalars $\sigma ^{2}$ and $\Theta $ are given by
(\ref{Lsigma2}) and (\ref{LTheta}) respectively. Using these expressions we
see that $\Sigma $ is constant and 
\begin{equation}
\Sigma =\frac{1-r}{1+2r}\, .  \label{LSigma}
\end{equation}%
Given that $\Sigma >0$ and $0<r<1$, we immediately deduce that $0<\Sigma <1,$
in accord with ${\cal R}<0$, in (\ref{LcR}). Thus, although isotropy ($%
\Sigma =0$) is not {\it asymptotically stable} (in the Lyapunov sense), it
is {\it stable} in the sense that any deviations from isotropy never diverge~%
\cite{B1, BTclosed, BS2} and $\Sigma $ tends to a constant at large times. Note
that when we set $r\rightarrow 1$ the $\Sigma $-parameter approaches zero
and the expansion becomes isotropic (i.e. the Milne universe).

We can therefore also write the   power-law evolution of the average scale
factor as $a\propto t^{1/(1+2\Sigma )}$. Thus, in the absence of any shear
anisotropy we have $a\propto t$, as in the Milne universe. For maximum shear
anisotropy as $r\rightarrow 0$, we obtain the familiar scale-factor
evolution characteristic of the Kasner vacuum or kinetic-dominated scalar
field solution \ref{sec:perts} (i.e.~$a\propto t^{1/3}$).

The trace of the 3-Ricci tensor ${\cal R}_{\alpha \beta }$ associated with
the surfaces of constant time is 
\begin{equation}
{\cal R}=-\frac{k^{2}(1-f^{2})^{2}}{2f^{2}t^{2}}-\frac{6r^{2}}{t^{2}}=-\frac{%
2r(1+2r)}{t^{2}}\,<0.  \label{LcR2}
\end{equation}

Spatial curvature anisotropies are described via the symmetric and
trace-free tensor ${\cal S}_{\alpha \beta }$. The only non-zero components
of ${\cal S}_{\alpha \beta }$ are 
\begin{equation}
{\cal S}_{11}=-\frac{4r(1-r)}{3t^{2}}\,,\hspace{5mm}{\cal S}_{22}=\frac{%
k^{2}(1-f^{2})(2+f^{2})}{3f^{2}t^{2}}\,,\hspace{5mm}{\cal S}_{33}=-\frac{%
k^{2}(1-f^{2})(1+2f^{2})}{3f^{2}t^{2}}  \label{LcS1}
\end{equation}%
and 
\begin{equation}
{\cal S}_{23}=-\frac{kr(1-f^{2})}{ft^{2}}\,.  \label{LcS2}
\end{equation}%
According to (\ref{LcR2})-(\ref{LcS2}), the spatial curvature of the model
vanishes at the maximum shear limit, namely as $r\rightarrow 0$, hence the
approach to the Kasner expansion rate. When $r$ $\rightarrow 1$, only the
isotropic part of ${\cal R}_{\alpha \beta }$ survives as $k^{2}(1-f^{2})=0$
as $r\rightarrow 0$ or $1$.

\end{appendix}

\bibliographystyle{JHEP}
\bibliography{refs}

\end{document}